\documentclass[twocolumn,10pt, twoside]{IEEEtran}
\IEEEoverridecommandlockouts

\usepackage[utf8]{inputenc}
\usepackage{nccmath}
\usepackage[english]{babel}
\usepackage{float}
\usepackage{wrapfig}
\usepackage{subfigure}
\usepackage{graphicx}
\usepackage[export]{adjustbox}
\graphicspath{ {images/} }
\usepackage{amsmath}
\usepackage{bm}
\usepackage{algorithmic}
\usepackage{array}
\usepackage{stfloats}
\usepackage[nolist]{acronym}
\usepackage{booktabs}
\usepackage{siunitx}
\usepackage[flushleft]{threeparttable}
\usepackage{xcolor}
\usepackage{amsthm, amsfonts, color}
\usepackage{verbatim}
\usepackage{amssymb}
\usepackage{listings}
\usepackage[noadjust]{cite}
\usepackage{tikz}
\usepackage{pgfplots}
\usepackage{makecell}
\usepackage{colortbl}
\usepackage{mathtools}
\usepackage{multirow} 
\usepackage{bbm}
\usepackage{LVcolours}
\usepackage{winsnotation}
\usepackage{siunitx}
\usepackage{upgreek}
\usepackage{hyperref}

\newcommand{\pnt}{\mathrm{pnt}}
\newcommand{\equ}{\mathrm{eq}}
\newcommand{\SNR}{\mathsf{SNR}}
\newcommand{\atm}{\mathrm{atm}}
\newcommand{\opt}{\mathrm{opt}}
\newcommand{\out}{\mathrm{out}}
\newcommand{\eff}{\mathrm{eff}}
\newcommand{\ave}{\mathrm{ave}}
\newcommand{\EIRP}{\mathsf{EIRP}}
\newcommand{\extra}{\mathrm{extra}}
\newcommand{\syst}{\mathrm{syst}}
\newcommand{\elevationangle}{\varphi}

\newcommand{\dB}{\mathrm{dB}}
\newcommand{\dBi}{\mathrm{dBi}}
\newcommand{\rmt}t%{\mathrm{t}}
\newcommand{\rmr}r%{\mathrm{r}}
\newcommand{\Opt}{\mathrm{opt}}
\newcommand{\las}{\mathrm{las}}

\newcolumntype{C}[1]{>{\centering\arraybackslash}p{#1}}

\title{Two-Leg Deep Space Relay Architectures: Performance, Challenges, and Perspectives}
\author{Dario Modenini, \IEEEmembership{Member, IEEE}, Alfredo Locarini, Lorenzo Valentini, \IEEEmembership{Student~Member,~IEEE},\\ Alberto Faedi, Paolo Tortora, \IEEEmembership{Senior Member,~IEEE},
Davide Rovelli, Nicol{\`o} Mazzali, \IEEEmembership{Member,~IEEE},\\ Marco Chiani, \IEEEmembership{Fellow,~IEEE}, Enrico Paolini, \IEEEmembership{Senior Member, IEEE}
\thanks{This work was supported by the European Space Research and Technology Centre (ESA/ESTEC) under contract 4000132053/20/NL/FE.}
\thanks{L. Valentini, A. Faedi, E. Paolini, M. Chiani, are with CNIT, DEI, University of Bologna, via Dell'Universit{\`a} 50, Cesena, Italy. Email: \{lorenzo.valentini13,alberto.faedi6,e.paolini,marco.chiani\}@unibo.it.}
\thanks{D. Modenini and P. Tortora are with DIN and CIRI Aerospace, University of Bologna, via Fontanelle 40, Forl{\`i}, Italy. Email: \texttt{\{dario.modenini,paolo.tortora\}@unibo.it}.}
\thanks{A. Locarini is with CIRI Aerospace, University of Bologna, via Fontanelle 40, Forl{\`i}, Italy. Email: \texttt{alfredo.locarini@unibo.it}.}
\thanks{Davide Rovelli is with the European Space Research and Technology Centre, Noordwijk, The Netherlands. Email: davide.rovelli@esa.int.}
\thanks{Nicol{\`o} Mazzali is with Modis for ESA, European Space Research and Technology Centre, Noordwijk, The Netherlands. Email: nicolo.mazzali@esa.int.}
\thanks{Opinions, interpretations, recommendations, and conclusions presented in this paper are those of the authors and are not necessarily endorsed by ESA.}
}

\begin{document}

\begin{acronym}
\acro{AWGN}{additive white Gaussian noise}
\acro{BER}{bit error rate}
\acro{CCSDS}{Consultative Committee for Space Data Systems}
\acro{DSA}{deep-space antenna}
\acro{EHF}{extremely high frequencie}
\acro{EIRP}{equivalent isotropically radiated power}
\acro{ESA}{European Space Agency}
\acro{FER}{frame error rate}
\acro{FOV}{field of view}
\acro{G/S}{ground station}
\acro{GEO}{Geostationary Earth Orbit}
\acro{i.i.d.}{independent and identically distributed}
\acro{ITU-R}{International Telecommunication Union--Radiocommunication}
\acro{LEO}{Low Earth Orbit}
\acro{LOS}{line of sight}
\acro{PPM}{pulse position modulation}
\acro{RF}{radio frequency}
\acro{S/C}{spacecraft}
\acro{SCPPM}{serially concatenated pulse position modulation}
\acro{SDP}{Sun - data relay - deep space probe}
\acro{SPD}{Sun - deep space probe - data relay}
\acro{TM}{telemetry}
\acro{TC}{telecommand}
\end{acronym}

\maketitle
\begin{abstract}
In this paper, architectures for interplanetary communications that feature the use of a data relay are investigated. In the considered ``two-leg'' architecture, a  spacecraft orbiting the Earth,  or in orbit at a Lagrange point, receives data from a deep space probe (leg-1)  and relays them towards ground (leg-2). Different wireless technologies for the interplanetary link, namely, radio frequencies above the Ka band and optical frequencies, are considered. Moreover, the cases of transparent and regenerative relaying as well as different different orbital configurations are addressed, offering a thorough analysis of such systems from different viewpoints. Results show that, under certain constraints in terms of pointing accuracy and onboard antenna size, the adoption of a two-leg architecture can achieve the data rates supported by direct space-to-Earth link configurations with remarkably smaller ground station antennas.
\end{abstract}

\section{Introduction}

\IEEEPARstart{T}{he} use of extremely high \ac{RF} bands or optical frequencies in wireless digital communication systems is known to potentially allow achieving very high data rates, compared with the ones achievable at lower frequencies, for the same error rate performance. 
Such extremely high frequency bands are, however, seldom employed in deep space communication links due to their vulnerability to atmospheric impairments. 
They may, nevertheless, provide several advantages in deep space \ac{TM} and \ac{TC} links in the framework of a “two-leg” relay architecture. 
Accordingly, a \ac{S/C} orbiting the Earth or in orbit at a Lagrange point would receive \ac{TM} data from a deep space probe and would relay them to the \ac{G/S}. 
The deep space to relay link, not affected by the Earth atmosphere, may take advantage of an extremely high frequency band, e.g., frequencies between Ka-band ones and $75$ GHz (which include the Q/V band), or the optical band, while the second link may use a more classical RF band, such as the K-band (for a near-Earth relay) or the X/Ka band (for a relay in a Lagrange point), benefiting from a shorter distance to the ground. 
The objective of this paper is to provide a thorough analysis of two-leg deep space architectures, with reference to \ac{TM} links, assessing their potential advantages with respect to a classical direct link architecture.  
Emphasis is put on scenarios in which the deep space probe is orbiting another planet of the solar system, so that the first link of the two-leg architecture, from the deep space probe to the relay, is an interplanetary one. 

Space data relaying has been attracting interest for a long time, being first envisaged by NASA in the late 60s \cite{Stampfl1970:Tracking}. 
More than a decade later, in 1983, the first tracking and data relay satellite service ever became operational, with the aim of providing near continuous communications and tracking services to  \ac{LEO} spacecrafts, launch vehicles, and suborbital platforms in general. 
Although several alternatives to offer data relay services to satellites orbiting in Earth proximity have been envisaged, mainly from national agencies, but also from commercial companies (e.g., \cite{Kopp2019:Utilizing,Hogie2015:TDRSS}), to date, only two successful relay satellite systems exist: (1) The Tracking and Data Relay Satellite System (TDRSS) and (2) The European Data Relay Satellite System (EDRSS) \cite{Witting2012:Status}. 

The reason behind considering LEO spacecrafts as the main use case for relay services lies in that, from a purely coverage point of view, users getting the maximum benefit are spacecrafts orbiting below  \ac{GEO} altitude. 
On the other hand, since deep space or planetary exploration missions can be tracked with a limited number of \acp{G/S} in most situations, they have received less attention so far as potential users of relay constellations. 
Nonetheless, a few studies are available, mainly from NASA, assessing the feasibility of communication relay satellites in GEO orbit for deep-space users, operating whether at radio \cite{Hunter1978:Orbiting}, or optical \cite{Wilson2003:Cost}, frequencies.
Recent studies involving relay satellites for deep space are primarily focused on optical links, \cite{Wittig2009:Data,Hemmati2011:Deep,Cesarone2011:Deep,Cornwell2017:NASA,Hurd2006:Exo}, discussing performance requirements, candidate configurations, and future trends. 
Proposed concepts include an application of EDRS for data relaying with near-Earth and deep space probes, called Data Relay for Moon (DROM), with possible extensions to Mars orbiters and next generation NASA GEO optical relay satellites, aimed at offering data rates up to $2.88\,\mathrm{Gbps}$ to Lunar users and data rates in the order of $100\,\mathrm{Mbps}$ to a deep space mission to Psyche asteroid \cite{Cornwell2017:NASA}.
To the best of the authors' knowledge, however, none of the existing studies provides a systematic analysis of deep-space data relay systems encompassing, at the same time, different planetary targets, data relay orbital configurations and frequency bands, a gap that this work aims to fill.

In this work, we investigate two-leg deep space relay architectures with the goal of assessing their potential advantages over classical direct \ac{RF} links for interplanetary communications. 
The analysis is carried out by considering (1) different wireless technologies for the interplanetary link, namely, radio frequencies above the Ka band and optical frequencies; (2) different relaying strategies, in particular transparent and regenerative relaying; (3) different orbital configurations, including the relay position and the type of target (inner or outer planet).
The number of degrees of freedom available to compare an architecture based on a direct link with a two-leg architecture is large and the problem may be tackled from several perspectives.
Concerning link performance analysis, in this paper we take the following approach: we assume that the two-leg system shall support on both links the same data rate supported by the direct-link system, with the same error probability, and we investigate the benefits of the two-leg architecture in terms of ground antenna size. 
As a main outcome of this work, we show that in some cases the ground antenna size can be considerably reduced, provided specific requirements in terms of \ac{S/C} pointing accuracy and onboard antenna size can be met.

This work is organized as follows. 
Section~\ref{prelim} clarifies the notation used throughout the paper and introduces preliminary elements such as the considered scenarios and the performance of direct links. 
In Section~\ref{orbital}, different alternatives for the orbit of the data relay \ac{S/C} are discussed and compared.
Section~\ref{sec:two_leg_framework} defines a framework for analysis of two-leg deep space relay systems, including the case of transparent and regenerative relay, and addressing the problem of pointing losses and optical link analysis. Performance results are presented in Section~\ref{sec:results}, while system engineering resources for an optical link are discussed in Section~\ref{sec:sys_res_optical}.
Finally, conclusions are drawn in Section~\ref{sec:conclusions_perspectives}, where a discussion about perspectives for deep space two-leg systems is also included.
A subset of the results, about pointing losses and optical link performance, appeared in the conference paper \cite{ValFae:21}.
With respect to \cite{ValFae:21} this paper includes results on orbital configurations, two-leg systems based on extremely high frequencies, as well as the system engineering resources analysis.

\section{Preliminaries and Notation}\label{prelim}

\subsection{Notation}\label{subsec:notation}

When referring to a two-leg architecture, we use the subscript ``1'' to refer to the first link, i.e., the interplanetary link between the deep space probe and the relay. 
This link is hereafter also referred to as ``leg-1''.
Similarly, the subscript ``2'' refers to the second link, i.e., the one between the relay and the \ac{G/S}.
We often use the expression ``leg-2'' to refer to this second link.
Then, for example, $f_1$ and $f_2$ denote the leg-1 and leg-2 carrier frequencies, respectively.
With reference to any link (either from space to ground or from space to space), $P_t$ and $P_r$ denote the transmit and receive power, $G_t=G_t(f)$ and $G_r=G_r(f)$ the transmit and receive antenna gains, $A_t$ and $A_r$ the \ac{S/C} transmit loss and the (\ac{G/S} or \ac{S/C}) receive implementation loss, and $r$ the range (distance between the transmitter and the receiver), $A_{\pnt, r} (f) = A_{\pnt, r}$ and $A_{\pnt, r} (f) = A_{\pnt, r}$ the attenuation due to pointing miss-match. We also define the inverse of the pointing attenuation as  $L_{\pnt} = 1 / A_{\pnt}$ referred through the paper as pointing loss. 
Concerning antennas, $d_t$ and $d_r$ denote the transmit and receive antenna sizes (i.e., diameters), while $\eta_t$ and $\eta_r$ are the corresponding efficiencies.
For residual carrier modulation, the portion of the total receive power apportioned to data is indicated as $P_d$. 
When analyzing a two-leg architecture where non-optical frequencies are used in leg-1, we are mostly interested in values of $f_1$ belonging to the Q and V bands. For these frequencies we often adopt the nomenclature \acp{EHF}, although in principle the \ac{EHF} band is much wider, spanning from $30\,\mathrm{GHz}$ to $300\,\mathrm{GHz}$.

Throughout the paper we denote by $A_{\atm} = A_{\atm} (f,\elevationangle,p)$ the atmospheric attenuation of a space-to-ground link, where $f$ is the frequency, $\elevationangle$ is the \ac{G/S} antenna elevation angle, and $p$ is the percentage of the time in which the link is unavailable.
This latter parameter is directly related to the link availability: For example, $p=5$ corresponds to $95\%$ availability.
In all cases, the atmospheric attenuation is computed as $A_{\mathrm{atm}}=A_{\mathrm{rain}}+A_{\mathrm{gas}}+A_{\mathrm{clouds}}$, i.e., by including attenuation caused by rain, absorption from gases (e.g., oxygen and water vapor), and small droplets (clouds and fog), according to \ac{ITU-R} recommendations, e.g., \cite{ITU-R2019:P.618-13,ITU-R2019:P.676-12,ITU-R2019:P.840-8}. 
The atmospheric attenuation $A_{\atm} (f,\elevationangle,p)$ actually depends on a number of additional site-specific parameters at the \ac{G/S}, which include the height above mean sea level, the point rainfall rate for $0.01\%$ of an average year, the atmospheric pressure, the temperature, the water vapour density, and the total columnar content of liquid water reduced to a temperature of $273.15\,\mathrm{K}$.
All of these parameters have been estimated for the location of \ac{ESA} \ac{DSA} 1, located in New Norcia, Australia, using the data made available by \ac{ITU-R}.
We also acknowledge that, following ESA Alphasat mission \cite{Paraboni:Mission}, data from several propagation experiments between Alphasat's Aldo Paraboni payload and different ground stations in Europe are made available in the literature, covering two frequencies at Ka and Q bands ($19.7\,\mathrm{GHz}$ and $39.4\,\mathrm{GHz}$, respectively) \cite{Paraboni:France, Paraboni:Italy,Paraboni:Slovenia,Paraboni:Austria}. 
These experiments report in general fairly good agreement with \ac{ITU-R} models, although with different levels of matching.
Since, to the best of authors' knowledge, a new model based on the above-mentioned experiments has not yet been consolidated, in this work the choice has been made to stick to the current \ac{ITU-R} models for atmospheric attenuation computation. 

When analyzing the mutual geometry between the deep space probe, the Sun and the data relay \ac{S/C}, the first will be referred as the ``probe'', identified with the letter ``P'', and the last will be identified with the letter ``D'', so that the angular separation between the Sun ``S'' vector and the transmitter-receiver line of sight will be denoted as \ac{SDP} angle or \ac{SPD} angle depending on whether we are considering the \ac{FOV} of the data relay spacecraft or that of the deep space probe, respectively.

Finally, throughout the whole paper we denote by $B_r$ the data rate, i.e., the information bit rate at the input of the channel encoder. The energy per information bit is $E_b$ while the one-sided noise power spectral density is $N_0 = k  T_{\syst}$, where $k=1.38 \cdot 10^{-23}\,\mathrm{J}/\mathrm{K}$ is the Boltzmann constant and $T_{\syst}$ is the receiver system noise temperature expressed in Kelvin.

\subsection{Reference Scenarios}

Deep space missions targeting a selection of inner (Venus) and outer (Mars, Uranus, and Neptune) Solar System planets are considered to establish some reference application cases. 
Emphasis is put, whenever possible, to the most recent missions, which are more representative of the state-of-the-art. 
For each central body considered as a possible target, a relevant scenario was derived for a classical direct space-to-Earth link, based on data from the surveyed missions involving orbiting spacecraft (i.e., excluding fly-by only ones). 
The key parameters adopted to characterize the deep space direct communication link are: transmitter-receiver maximum distance, downlink frequency band, transmit power, and antenna size. The reference scenarios for Venus, Mars, Uranus and Neptune are reported in Table~\ref{table:Venus_table}. Note that the range values reported in the table correspond to the worst-case range.

\begin{table*}[!t]
\begin{center}
\caption{Reference scenarios for missions to Venus, Mars, Uranus and Neptune}\label{table:Venus_table}
\footnotesize{
\begin{tabular}{ccccc}
\toprule
 {\textbf{Central Body}} & {\textbf{Tx–Rx Distance}} & {\textbf{Downlink Band}} & {\textbf{Downlink Tx Power}} & {\textbf{S/C Antenna Size}} \\
 \toprule
\textbf{Venus} &  259,594,256 km (1.735280 AU) & X-Band / Ka-Band &	65 W / 120 W & 1.3 m / 2.5 m\\
\midrule
\textbf{Mars} &  400,373,069 km (2.676328 AU) & X-Band &	65 W & 2.2 m\\
 \midrule
\textbf{Uranus} &  3,155,452,867 km (21.092899 AU) & X-Band & 65 W & 3 m\\
 \midrule
\textbf{Neptune} &  4,687,074,599 km (31.331158 AU) & X-Band / Ka-Band &	65 W / 100 W & 3 m\\
\bottomrule
\end{tabular}}
\end{center}
\end{table*}

%\EPcomment{ADD TABLES HERE}

\subsection{Performance of Classical Direct Links}\label{subsec:direct}

The performance of a classical direct (one-leg) space-to-Earth link can be evaluated through a standard link budget procedure.
Specifically, we have
\begin{align}
\frac{P_r}{N_0} = \frac{P_t\, G_t G_r }{  A_t  A_r A_{\atm}  A_{\pnt,t}  A_{\pnt,r}  \,k T_{\syst}} \, \frac{c^2}{(4 \pi f r)^2}
\end{align}
where $c$ is the light speed and the meaning of the other parameters involved in the expression has been clarified in Section~\ref{subsec:notation}. 
Moreover, the portion of the total received power apportioned to data, $P_d$, fulfills
\begin{align}
P_d/N_0 = \left\{ \begin{array}{ll}
P_r/N_0 & \text{suppressed carrier modulation} \\
   P_r/N_0  \sin^2 (\beta) & \text{residual carrier modulation} 
\end{array} \right.
\end{align}
where $\beta$ is the modulation index (for residual carrier modulation). Recommended residual carrier modulation types are PCM/PSK/PM and PCM/PM/Bi-$\phi$ \cite{ECSS2011:50-05C,Shihabi1994:Comparison}. The available $E_b/N_0$ at the \ac{G/S} is
\begin{align}\label{eq:EbN0}
\frac{E_b}{N_0} = \frac{P_d}{N_0 B_r}    
\end{align}
where, as mentioned above, $B_r$ is the data rate.
The $E_b / N_0$ margin can be computed as the difference between the available $E_b / N_0$ \eqref{eq:EbN0} and the parameter $(E_b /N_0)^*$, referred to as the required $E_b /N_0$ and representing the minimum $E_b /N_0$ value to achieve the target \ac{BER} or \ac{FER}. The value of $(E_b /N_0 )^*$ depends on the employed coding and modulation.

\begin{table}[!t]
\begin{center}
\caption{Achievable data rates for a direct X-band link, for different central bodies at maximum range.}\label{table:data_rates}
\footnotesize{
\begin{tabular}{cc}
\toprule
 {\textbf{Central Body}} & {\textbf{Data rate}} \\
 \toprule
Venus & $100\mathrm{kbps}$\\
\midrule
Mars &  $120\mathrm{kbps}$\\
 \midrule
Uranus & $3.15\,\mathrm{kbps}$\\
 \midrule
Neptune &  $1.20\,\mathrm{kbps}$\\
\bottomrule
\end{tabular}}
\end{center}
\end{table}

Table~\ref{table:data_rates} summarizes the achievable data rate values in the four considered scenarios for a direct X-band link at maximum range.
These values have been obtained via standard link budget assuming the parameters reported in Table~\ref{table:Venus_table}, and targeting \ac{ESA} \ac{DSA}-1, characterized by $d_r = 35\,\mathrm{m}$ and $G/T_{\syst} = 50.1\,\mathrm{dB}$ \cite{EFM:08,besso2015:Present}.
The transmit antenna efficiency and the transmit and receive losses are set to $\eta_t=0.7$, $A_t=1.5\,\mathrm{dB}$, and $A_r=0.5\,\mathrm{dB}$ in all scenarios.
The X-band atmospheric attenuation, computed for $95\%$ availability and $\elevationangle=10$° elevation angle (commonly employed values for the X-band) turns to be $A_{\atm}=0.5\,\mathrm{dB}$, while pointing losses turn to be negligible.
In compliance with recommendations, the modulation schemes are residual carrier ones while the employed channel coding schemes are the rate-$1/4$ $(28560,7136)$ turbo code for Venus and Mars and the rate-$1/6$ $(42840,7136)$ turbo code for Uranus and Neptune.
The $E_b/N_0$ margin is $4\,\mathrm{dB}$ is all cases.

\section{Two-Leg System Orbital Configurations} \label{orbital}

We consider different two-leg orbital geometries, assuming the relay to be placed whether on LEO, GEO or at a lagrangian point of the Earth-Sun system, and the deep space probe orbiting at an inner or outer planet. The resulting configurations are compared according to their complexity, occultation/conjunction time, and range.
The typical distance between the relay S/C and an hypothetical ground station, involved in each scenario, is included in Table~\ref{table:GS_DR_range}

\begin{table}[!t]
\begin{center}
\caption{Ground station - Data Relay S/C range}\label{table:GS_DR_range}
\footnotesize{
\begin{tabular}{cc}
\toprule
 {\textbf{Data Relay S/C orbit}} & {\textbf{Leg-2 range}} \\
 \toprule
LEO & $1-3 \cdot 10^{3}\mathrm{km}$\\
\midrule
GEO & $3-4 \cdot 10^{4}\mathrm{km}$\\
 \midrule
L1/L2 & $\sim1.5 \cdot 10^{6}\mathrm{km}$\\
 \midrule
L4/L5 & $\sim1.5 \cdot 10^{8}\mathrm{km}$\\
\bottomrule
\end{tabular}}
\end{center}
\end{table}

\subsection{Relay Spacecraft in LEO}

The exploitation of the near-Earth space for data relay to support deep-space missions provides some advantages in terms of reduction of the orbital deployment costs, as they would benefit of the wide offer of launches targeting LEO heights. 
Moreover, a LEO data buffer enables the possibility to use any of the numerous small stations commonly employed in near-Earth missions, spread at a very wide range of latitudes and longitudes. 

However, to guarantee an adequate geometric visibility between the relay and the deep space probe, which may be requested to be nearly $100\%$ during the most critical mission phases, multiple LEO relays would be required. 
As an additional drawback, a deep space observer would experience an angular separation between the Earth and the relay spacecraft so small that the Earth would act as a continuous source of noise for an hypothetical optical receiver.

\subsection{Relay Spacecraft in GEO}

Most of the advantages and disadvantages for a LEO relay \ac{S/C} apply also to the GEO scenario, the main difference being the drastic reduction of Earth occultation at the geosynchronous distance. 
Although the range between the transmitter and the receiver in leg-2 is larger than for the LEO case, it still allows using smaller stations than the ones employed in deep space missions, with the additional benefit of a simpler ground segment architecture, by selecting a station which is always in view of the GEO satellite. 
On the other hand, the space segment architecture would result more complex than for the LEO case, because of the need for an on-board propulsion system for orbit deployment and maintenance.

\subsection{Relay Spacecraft in a Lagrange Point}
L4 and L5 (equilateral) Lagrange points are the more promising candidates for data relay, being the only stable among the five equilibrium points of the Sun-Earth system. 
They are placed on the same trajectory covered by the Earth around the Sun, preceding and following the Earth, respectively, with a phase difference of approximately $60$°. L1/L2 Lagrange points are also considered, despite the additional complexity resulting from the expected increased station-keeping effort make them less attractive.
Although placing a relay on L4 or L5 lead to the highest range ($1\,\mathrm{AU}$) for leg-2, the high angular separation from the Earth would allow for a favourable observation geometry from both inner and outer planets, when either the Earth or the L4 (or L5) point is in solar conjunction with the target. 

\subsection{Geometric Visibility Analysis}

\begin{figure}
    \centering
    \includegraphics[width=0.9\columnwidth]{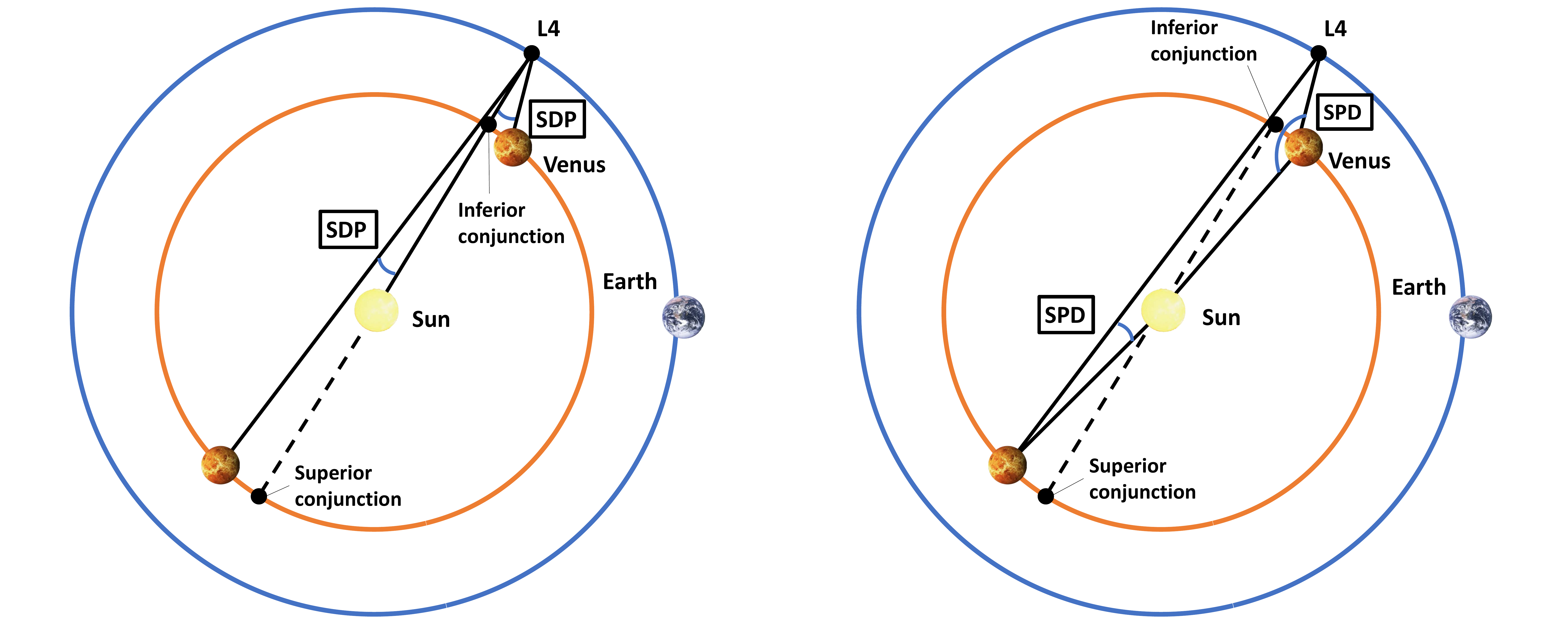}
    \includegraphics[width=0.9\columnwidth]{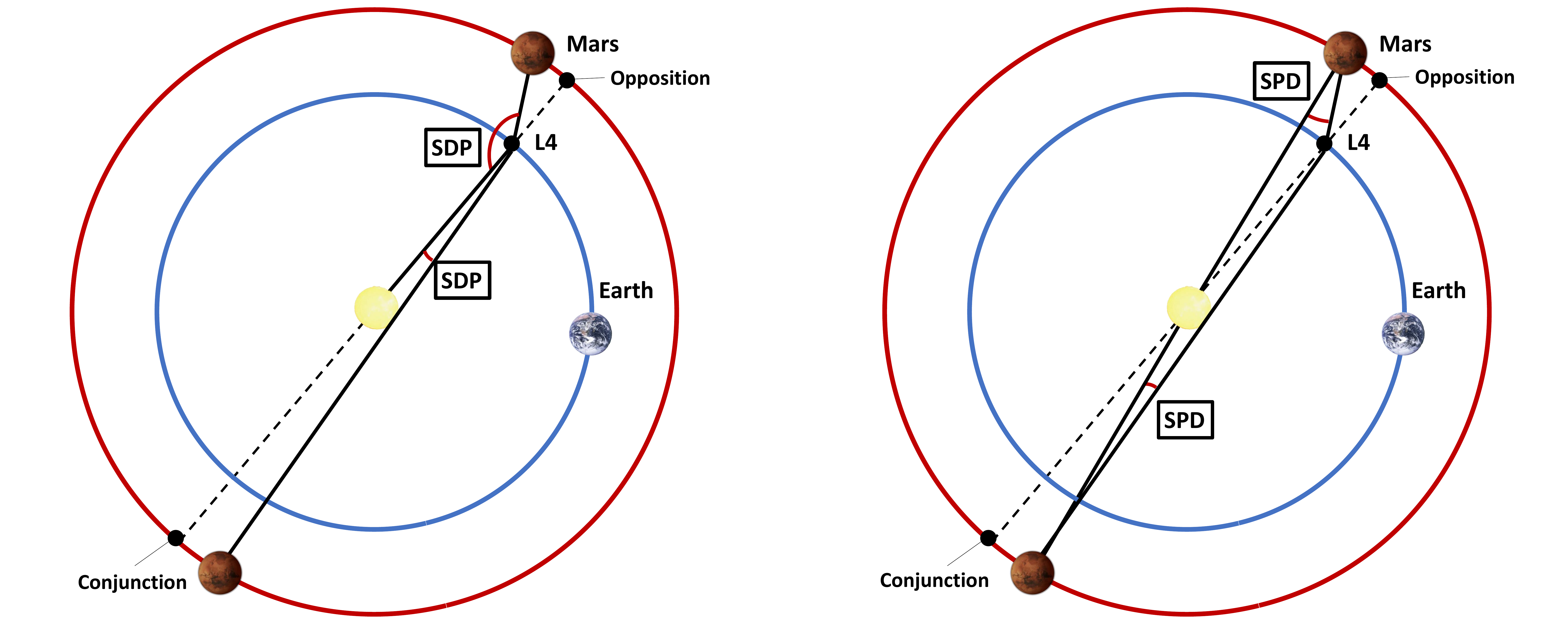}
    \caption{Graphical visualization of the \ac{SDP} and \ac{SPD} angles for a deep space probe to inner and outer planets}
    \label{figSDPSPD}
\end{figure}

The relative geometry between the Sun, the relay \ac{S/C} and the target planet will affect the percentage of link outage due to the possible presence of the Sun in (or close to) the line-of-sight between the transmitter and the receiver. This is especially true when an optical link is chosen for leg-1, as Sun radiation is by far the largest noise source which would affect such a link.
To this end, an analysis has been performed which, starting from the mutual orbital configurations, predicts the temporal evolution of the \ac{SDP}/\ac{SPD} angles for each planetary target of interest over the next 100 years. 
Fig.~\ref{figSDPSPD} depicts the \ac{SDP} and \ac{SPD} geometries for the cases of a mission to both an inner and an an outer planet.

We compute the link outage percentage as the amount of time, over the total mission lifetime, at which the \ac{SDP}/\ac{SPD} angle is below a certain value, resulting in the impossibility to establish the link.
Since such a threshold is (optical) hardware dependent, the comparison among different configurations should include computation of i) the outage percentage corresponding to different cut-off values of \ac{SDP}/\ac{SPD} angles, and ii) the minimum \ac{SDP}/\ac{SPD} angle that the optical instrument shall be capable to operate at, for an assumed percentage of link availability loss (a value of $10\%$ was adopted for this work).

In the framework of the \ac{SDP}/\ac{SPD} analysis, a data relay \ac{S/C} placed at the Lagrange points L4 and L5 or in LEO/GEO orbit around the Earth are treated as a single scenario, as they all move within the same orbital plane and at the same distance from the Sun. This means that the “relay–Sun” vector presents the same temporal evolution (with a phase offset) for any of the L4, L5 or Earth orbiting scenario, i.e., the same \ac{SDP} / \ac{SPD} dynamics.
Since placing a relay in L1/L2 results to only slight differences in the numerical values with respect to the L4/L5/LEO/GEO case, no explicit distinction will be made when presenting the data, for the sake of brevity. 

\begin{table}[!t]
\begin{center}
\caption{Outage percentage assuming \ac{SDP} thresholds equal to 1°, 3°, 10°, 40°.}\label{table:OutageSDP_perc}
\footnotesize{
\begin{tabular}{ccccc}
\toprule
 {\textbf{Outage SDP}} & {\textbf{Venus}} & {\textbf{Mars}} & {\textbf{Uranus}} & {\textbf{Neptune}} \\
 \toprule
\textbf{40°} &  69.67\% & 35.92\% &	23.41\% & 23.37\%\\
\midrule
\textbf{10°} &  14.95\% & 9.08\% &	5.85\% & 5.82\%\\
 \midrule
\textbf{3°} &  3.84\% & 2.65\% & 1.76\% & 1.63\%\\
 \midrule
\textbf{1°} &  0.25\% & 0.48\% & 0.45\%	& 0.23\%\\
\bottomrule
\end{tabular}}
\end{center}
\end{table}

\begin{table}[!t]
\begin{center}
\caption{Outage percentage assuming \ac{SPD} thresholds equal to 1°, 3°, 10°, 40°.}\label{table:OutageSPD_perc}
\footnotesize{
\begin{tabular}{ccccc}
\toprule
 {\textbf{Outage \ac{SPD}}} & {\textbf{Venus}} & {\textbf{Mars}} & {\textbf{Uranus}} & {\textbf{Neptune}} \\
 \toprule
\textbf{40°} &  37.75\% & 87.27\% & 100\% & 100\%\\
\midrule
\textbf{10°} &  9.55\% & 16.89\% & 100\% & 100\%\\
 \midrule
\textbf{3°} &  2.55\% & 4.64\% & 94.45\% & 100\%\\
 \midrule
\textbf{1°} &  0.53\% & 1.24\% & 21.95\% & 35.04\%\\
\bottomrule
\end{tabular}}
\end{center}
\end{table}

Results are illustrated by means of tables. Table~\ref{table:OutageSDP_perc} and Table~\ref{table:OutageSPD_perc} display the outage percentages for the cases when \ac{SDP} (or \ac{SPD}) is lower than $1$°, $3$°, $10$°, and $40$°. 
Moreover, Table~\ref{table:MinAngle} displays the value of the minimum \ac{SDP} and \ac{SPD} angles that the data relay and deep space probe communication equipment, respectively, must be able to operate with when considering a maximum loss of link availability due to the solar noise of $10\%$. 
Results indicate that, at Venus (Mars), the optical equipment must be designed to operate with \ac{SDP}/\ac{SPD} values down to $7.1$° ($11$°) for the relay \ac{S/C} and $10.6$° ($6.01$°) for the deep space probe. 
Deep space probes at Uranus and Neptune must instead cope with more severe optical design requirements in terms of straylight rejection capabilities and narrower \ac{FOV}, since lower \ac{SPD} values are involved.

\begin{table}[!t]
\begin{center}
\caption{Min SDP associated to Link availability loss of 10\%.}\label{table:MinAngle}
\footnotesize{
\begin{tabular}{ccccc}
\toprule
 {\textbf{}} & {\textbf{Venus}} & {\textbf{Mars}} & {\textbf{Uranus}} & {\textbf{Neptune}} \\
 \toprule
\textbf{Outage \ac{SDP}} &  7.1° & 11.0°	& 17.0°	& 17.1°\\
\midrule
\textbf{Outage \ac{SPD}} &  10.46° & 6.01° & 0.46° & 0.30°\\
\bottomrule
\end{tabular}}
\end{center}
\end{table}

A relay \ac{S/C} in L4/L5 would benefit from a larger angular separation from the Earth: Table~\ref{table:AngleShift} shows the minimum values of \ac{SDP}/\ac{SPD} which result from assuming to switch between an optical communication to the L4/L5 relay and a direct RF link with Earth depending on which of the two geometries is more advantageous when approaching conjunction.
When either L4 or L5 are at low \ac{SDP} angles, ground-based deep space antennas on the Earth may be used to send telecommands to the probe. The same consideration applies also for the data downlink from a deep space probe at Venus and Mars, while at Uranus or Neptune the \ac{SPD} would still be very low even considering the alternative direct link to the Earth.
To be noted that such an advantage may only be exploited under the assumption that the deep space probe is equipped with a redundant telecommunication segment, one optical and one RF segment, as optical link is not considered a suitable solution for direct communication with Earth, because of its high sensitivity to the atmospheric conditions.

\begin{table}[!t]
\begin{center}
\caption{Min \ac{SDP}/\ac{SPD} angles achievable with a combined L5 + Earth communication architecture.}\label{table:AngleShift}
\footnotesize{
\begin{tabular}{ccccc}
\toprule
 {\textbf{}} & {\textbf{Venus}} & {\textbf{Mars}} & {\textbf{Uranus}} & {\textbf{Neptune}} \\
 \toprule
\textbf{Min \ac{SDP} L5+Earth} & 12.39° & 17.51° & 28.55° & 29.08°\\
\midrule
\textbf{Min \ac{SPD} L5+Earth} & 17.33° & 11.19° & 1.37° & 0.92°\\
\bottomrule
\end{tabular}}
\end{center}
\end{table}

\section{Two-Leg System Link Analysis Framework}\label{sec:two_leg_framework} 

Following the existing literature, e.g., \cite{Matricciani2005:Deep,Matricciani2009:Optimum} we distinguish between the two cases of transparent relay and regenerative one. 
A transparent relay performs signal amplification and carrier frequency conversion, without attempting any signal demodulation and channel decoding operation. 
Conversely, a regenerative relay demodulates and decodes the signal incoming from the spacecraft, re-encodes and re-modulates the data, and re-transmits them to the G/S, possibly using a carrier frequency different from the one of the received signal. 
In case of a transparent relay, the employed coding and modulation scheme is the same in both legs; in contrast, regenerative relays allow using different coding and modulation options and also different wireless technologies, e.g., optical and RF, for the two legs, but require data buffering capabilities.
A general end-to-end signal-to-noise ratio (SNR) analysis of a multi-hop communication system based on transparent relaying was developed in \cite{Hasna2003:Outage}.

\subsection{Transparent Relay}\label{subsec:transparent_framework}

We consider the case in which \ac{RF} signals are transmitted both links (hence, no optical technology is used). The signal power received by the relay in leg-1 is given by
\begin{align}
P_{r,1} =  \frac{ P_{t,1} G_{t,1} G_{r,1}  }{  A_{t,1}  A_{r,1}  A_{\pnt,t,1}  A_{\pnt,r,1} } \, \frac{c^2}{(4 \pi f_1  r_1 )^2} 
\end{align}
where we note the absence of atmospheric attenuation due to the relay being out of the Earth atmosphere. 
Denoting by $B_{\equ,1}$ the equivalent noise bandwidth of the receiver onboard the relay, the thermal noise power impairing the signal received in leg-1 is $N_1 = N_{0,1} B_{\equ,1} = k T_{\syst,1} B_{\equ,1}$. 
The SNR at the relay is therefore $\SNR_1 = P_{r,1} / N_1$. Since both signal and noise are amplified and forwarded by the relay, the power transmitted by the relay in leg-2 may be expressed as
\begin{align}
P_{t,2} = P_{t,2}^{(s)} + P_{t,2}^{(n)}
\end{align}
where $P_{t,2}^{(s)}$ is the amplified signal power and $P_{t,2}^{(n)}$ is the amplified noise power. Letting $P_{r,2}$ be the total power received on ground, contributed by both the signal and the noise forwarded by relay, the SNR at the G/S may therefore be expressed as
\begin{align}
\SNR_2 = \frac{P_{r,2}^{(s)}}{P_{r,2}^{(n)}+N_2} = \frac{P_{r,2}  \frac{P_{r,1}}{P_{r,1}+N_1}}{P_{r,2} \frac{N_1}{P_{r,1}+N_1}+N_2}
\end{align}
in which $N_2=N_{0,2} B_{\equ,2} = k T_{\syst,2} B_{\equ,2}$. Simple algebraic manipulation yields
\begin{align}
(\SNR_2)^{-1} = \frac{N_1}{P_{r,1}} + \frac{N_2}{P_{r,2}}  \left( 1+\frac{N_1}{P_{r,1}} \right)
\end{align}
where
\begin{align}
\frac{N_1}{P_{r,1}} = \frac{ A_{t,1}  A_{r,1}  A_{\pnt,t,1}  A_{\pnt,r,1}  N_1}{P_{t,1} G_{t,1} G_{r,1}  } \, \frac{(4 \pi f_1  r_1 )^2}{c^2}
\end{align}
and
\begin{align}
\frac{N_2}{P_{r,2}} = \frac{ A_{t,2}  A_{r,2}  A_{\pnt,t,2}  A_{\pnt,r,2} A_{\atm, 2} N_2}{P_{t,2} G_{t,2} G_{r,2}  } \, \frac{(4 \pi f_2  r_2 )^2}{c^2}
\end{align}
Next, we define an equivalent one-sided noise power spectral density $N_0$ as 
\begin{align}
N_0 = \frac{P_{r,2}^{(n)}+N_2 }{B_{\equ,2}}
\end{align}
which allows us writing
\begin{align}\label{eq:transparent_SN0}
\frac{P_{r,2}^{(s)}}{N_0} &= \SNR_2 B_{\equ,2} = \left[ \frac{N_1}{P_{r,1}} + \frac{N_2}{P_{r,2}} \left(1 + \frac{N_1}{P_{r,1}} \right) \right]^{-1} B_{\equ,2} \, .
\end{align}
The results presented in Section~\ref{sec:results} about two-leg relay systems with a transparent relay will mainly rely on the parameter $P_{r,2}^{(s)}/N_0$.

\subsection{Regenerative Relay}
\label{subsec:PeTot}
In the regenerative case, we do not impose use of \ac{RF} signals in leg-1, which may be either an \ac{EHF} or an optical link. 
Leg-2, instead, is assumed to be an \ac{RF} link. 
Denoting by $E_1$ and $E_2$ the decoding error events in the relay and in the \ac{G/S}, respectively, we have an error in the two-leg regenerative configuration when an  $E_2$ event occurs. We can develop the error probability as
\begin{align}\label{eq:regenerative_p_error}
\Pr(E_2) &= \Pr(E_2 | E_1) \Pr(E_1) + \Pr(E_2 | \bar{E}_1) \Pr(\bar{E}_1) \notag \\
& = \Pr(E_2 | E_1) \Pr(E_1) + \Pr(E_2 | \bar{E}_1) (1 - \Pr(E_1)) \, .
\end{align}
In the last row of \eqref{eq:regenerative_p_error}, $\Pr(E_2|E_1)$ is the probability to have a decoding error in the \ac{G/S} given that an error in the relay occurred. 
This probability is essentially $1$. 
Moreover, since the link budget in both links is designed to have small error probabilities (e.g., decoding error probability in the order of $10^{-4}$), we have $1-\Pr(E_1) \approx 1$. We then obtain, with very good approximation,
\begin{align}\label{eq:regenerative_pe_final}
\Pr(E_2) &\approx \Pr(E_1) + \Pr(E_2 | \bar{E}_1) = P_{\mathrm{e},1} + P_{\mathrm{e},2}
\end{align}
where $P_{\mathrm{e},1}=\Pr(E_1)$ and $P_{\mathrm{e},2}=\Pr(E_2|\bar{E}_1)$. Equation \eqref{eq:regenerative_pe_final} tells us that, in case of a regenerative relay, we can analyze the two links separately and design them under a constraint in the form
\begin{align}
P_{\mathrm{e},1} + P_{\mathrm{e},2} \leq P_{\mathrm{e}}^*
\end{align}
where $P_{\mathrm{e}}^*$ is the maximum tolerable decoding error probability.

\subsection{Pointing Losses}\label{subsec:Pointing}

The extremely narrow optical beams pose very tight requirements in terms of pointing accuracy. 
In fact, the miss-pointing losses between the transmit and the receive antennas cannot be neglected even in presence of a very accurate point-ahead calculation, due to mechanical noise generated by the vibrations and the satellite motions \cite{CheGar:89}. 
Moreover, despite the wider beam of \ac{RF} systems, \ac{EHF} links can also be affected by non-negligible pointing losses. 
This can be due, for example, to a lower accuracy of the \ac{RF} antenna pointing systems. 

Most available studies on pointing effects assume that only one antenna is affected by angular errors \cite{Vil:81}. 
For example, when one edge of the communication link is on ground, a very precise ground pointing can be assumed and, therefore, it is usual to consider the angular noise only on the \ac{S/C}.
In case of space-to-space links (such as leg-1 in a two-leg system), however, pointing effects should be carefully considered at both edges of the link. 
In the following we propose an approach to address the miss-pointing losses in situations where both terminals are affected by angular noise.
In addition, we introduce the concept of \emph{effective system gain} to optimize the antenna gain capturing pointing accuracy.

Pointing in a three-dimensional space is affected by both azimuth and elevation angular errors. 
Considering these two errors are \ac{i.i.d.} and Gaussian distributed with zero-mean, the resulting random angular error $\theta$ (i.e., the angle between the line of sight and the pointing direction) is Rayleigh distributed as \cite{CheGar:89, BarMec:85}
\begin{align}
\label{eq:ThetaRayleigh}
f_{\theta}(\theta) = \frac{\theta}{\sigma_{\theta}^{2}} \, e^{-\theta^2/2\sigma_{\theta}^{2}} , \quad \theta \geq 0
\end{align}
where $\sigma_{\theta}$ is the main angular noise parameter, hereafter referred to as the \emph{pointing accuracy}. 
We also need to assume a pointing loss model for both the receiver and the transmitter. 
For example, the Gaussian beam point loss model is given by \cite{CheGar:89}
\begin{align}
    L_{\pnt}(\theta) = e^{-G\theta^2}
    \label{eq:GaussianBeamLosses}
\end{align}
where $L_{\pnt} = 1/ A_{\pnt}$ is the inverse of the pointing attenuation referred through the paper as pointing loss, $G$ is the linear-scale antenna gain, and $\theta$ is the angular miss-pointing error. 
Another model is the circular aperture one where the pointing losses are instead given by \cite{CheGar:89,Orf:04}
\begin{align}
\label{eq:CircModel}
    L_{\pnt}(\theta) = \left( \frac{2 \, J_1(\sqrt{G}\theta)}{\sqrt{G}\theta} \right)^2
\end{align}
where $J_1(\cdot)$ is the Bessel function of the first kind of order 1. 

We may use two approaches to include pointing losses in the link analysis.
The first one is a deterministic approach which considers a maximum angular error $\theta_\mathrm{max}$ (based on the available information about the pointing system). Using this value it is possible to quantify the loss by employing, for example \eqref{eq:GaussianBeamLosses} or \eqref{eq:CircModel} models.  
Adopting a Gaussian beam model, for given $\theta_\mathrm{max}$ we compute the maximum pointing loss per antenna as
\begin{align}
L_{\pnt,\max} = e^{-G \theta^2_{\max}}
\end{align}
and we use it for link analysis.

We may also adopt a probabilistic approach. Accordingly, we define the outage probability as
\begin{align}
P_\mathrm{out} = \Prob{A_{\pnt,t}[\dB] + A_{\pnt,r}[\dB] > A_{\pnt}^{*} [\dB]}
\end{align}
where $A_{\pnt}[\dB]=-L_{\pnt}[\dB]$ is the pointing attenuation, the subscripts `$\rmt$' and `$\rmr$' refer to the transmitter and the receiver, respectively, and $A_{\pnt}^{*}[\dB]$ is the margin against the pointing attenuation. 
For given $P_\mathrm{out}$, $A_{\pnt}^{*}[\dB]$ represents the value of pointing attenuation to be included in the link budget. 
Under a Gaussian beam model and an angular error distribution \eqref{eq:ThetaRayleigh} we can derive $P_\mathrm{out}$ in closed-form as
\begin{align}
\label{eq:GauBeamPout}
    P_\mathrm{out} 
   &= \frac{G_\rmt\sigma_{\theta,\rmt}^{2} e^{-K /(2\sigma_{\theta,\rmt}^{2}G_\rmt)}}{G_\rmt\sigma_{\theta,\rmt}^{2}-G_\rmr\sigma_{\theta,\rmr}^{2}}\, + \frac{G_\rmr\sigma_{\theta,\rmr}^{2} e^{-K /(2\sigma_{\theta,\rmr}^{2}G_\rmr)}}{G_\rmr\sigma_{\theta,\rmr}^{2}-G_\rmt\sigma_{\theta,\rmt}^{2}}
\end{align}
where $K = \frac{\ln(10)}{10}\,A_{\pnt}^{*}[\dB]$. Considering a symmetrical system where $G_\rmt = G_\rmr = G$ and $\sigma_{\theta, \rmt} = \sigma_{\theta, \rmr} = \sigma_{\theta}$, the outage probability becomes
\begin{align}
\label{eq:GauBeamPoutGEqual}
P_\mathrm{out} = \left(1 + \frac{K}{2\sigma_{\theta}^{2}G}\right)\,e^{-K/2\sigma_{\theta}^{2}G}\,.
\end{align}
Different beam models or angular error distributions may require numerical evaluation.

\begin{figure}[t]
    \centering
    \resizebox{0.9\columnwidth}{!}{
    	\input{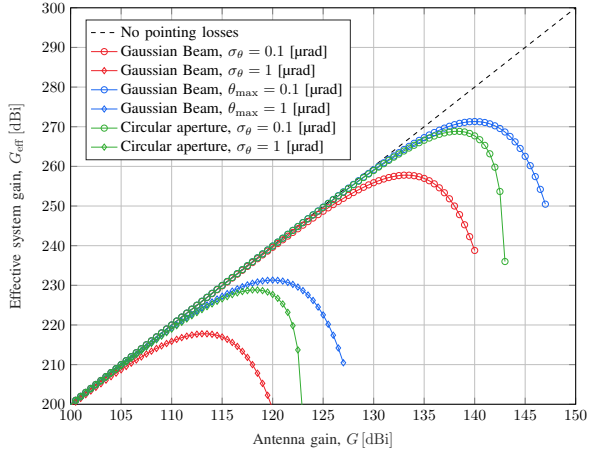}
    }
    \caption{Effective system gain \eqref{eq:Geff_G} versus the antenna gain $G=G_\rmt=G_\rmr$. Deterministic ($\theta_\mathrm{max}$-based) and probabilistic ($\sigma_\theta$-based) approaches; $P_\mathrm{out} = 5\%$ when probabilistic approach is employed; different values of the pointing accuracy parameters; Gaussian beam and circular aperture models.}
    \label{fig:EffectiveGain}
\end{figure}

Regardless of the approach used to estimate the pointing losses and the adopted model, we can define an ``effective system gain'' as
\begin{align}
G_\mathrm{eff} = (G_\rmt L_{\pnt,\rmt} ) \, (G_\rmr L_{\pnt,\rmr} )
\end{align}
i.e., as the product of the transmit and receive antenna gains, each multiplied by the corresponding pointing loss. 
Looking at the effective system gain reveals the existence of an optimal value for the antenna gains, beyond which the overall system performance degrades instead of improving. 
Considering a situation where the transmitter and the receiver are equipped with the same antennas, the effective system gain assumes the form
\begin{align}\label{eq:Geff_G}
G_\mathrm{eff}\,[\dBi] = 2\, G \,[\dBi] - A_{\pnt,\mathrm{tot}} \,[\dB]
\end{align}
where $A_{\pnt,\mathrm{tot}} \,[\dB]$ incorporates the total pointing attenuation. 
Sticking for simplicity to the case in which both antennas exhibit the same pointing accuracy parameter $\sigma_\theta$ or $\theta_\mathrm{max}$, we can numerically find $A_{\pnt,\mathrm{tot}} \,[\dB]$ (and therefore $G_\mathrm{eff}[\dBi]$) for any given gain $G[\dBi]$, for a particular value of $P_\mathrm{out}$ and $\sigma_\theta$ (or $\theta_\mathrm{max}$ if the deterministic approach is adopted). An example is shown in Fig.~\ref{fig:EffectiveGain} for different values of the pointing accuracy parameter $\sigma_\theta$ or $\theta_\mathrm{max}$ depending on the approach employed, different pointing loss models (Gaussian beam or circular aperture), and the outage probability $P_\mathrm{out} = 5\%$ for the probabilistic approach. 

\subsection{Optical Link Analysis}
\label{subsec:OptLinkAnalysis}

To date, the only coding and modulation scheme recommended for deep space optical telemetry is represented by \ac{SCPPM} \cite{CCSDS2019:142.0-B-1}. 
For peak and average power constraints typical of a deep-space link, restricting the modulation to \ac{PPM} is near-capacity achieving. 
The \ac{SCPPM} scheme is reviewed in Appendix~\ref{appendix:SCPPM}, along with some performance curves obtained via Monte Carlo simulation. 
Hereafter, we target use of \ac{SCPPM} in leg-1, when optical frequencies are used.

For the analysis of optical links we resort on optical link budget techniques \cite{Biswas2003:Mars,Biswas2020:Deep}, together with the above-introduced effective system gain $G_\mathrm{eff}$. 
Denote by $P_e^*$ the maximum tolerable value of decoding error probability, by $n_\mathrm{s}$ the actual received signal photon flux, expressed in dB phe/ns, and by $n_\mathrm{s}^*$ the minimum such number required to satisfy the $P_e \leq P_e^*$ requirement for a given coding and modulation scheme. 
The optical path loss attenuation, $A_{\opt}=A_{\opt} (\lambda,r)$, is given by $A_{\opt}  = ((4 \pi r) / \lambda)^2$ where $r$ is the range and $\lambda$ the wavelength.
Moreover, the transmit and receive optical antenna gains can be computed, similarly to the RF case\footnote{Conventionally, at optical frequencies antenna efficiencies are not incorporated in the gain.}, as
\begin{align}
G_t  = ((\pi D_t)/\lambda)^2 \qquad \mathrm{and} \qquad G_r  = ((\pi D_r)/\lambda)^2
\end{align}
respectively. 
For given $P_{\out}$ and $\sigma_\theta$ (probabilistic approach) or given $\theta_{\max}$ (deterministic approach) we can compute the antenna gains and the corresponding pointing losses, to maximize the effective system gain $G_{\eff}$. 
This is done by tuning the antenna diameter if $\lambda$ is fixed. 
In the following, we assume both $G_t$ and $G_r$ take their optimum values in terms of $G_{\mathrm{eff}}$ optimization.

The number of received signal photons $n_\mathrm{s}$ may be expressed as
\begin{align}
n_\mathrm{s}  = \EIRP\, G_r\,\eta_r  \, \frac{\lambda}{h c} \, A_{\opt}^{-1}
\end{align}
where the \ac{EIRP} is given by
\begin{align}
\EIRP = P_{\ave}\,G_t\,\eta_t\,L_{\pnt} \, L_{\extra}
\end{align}
and where $\eta_t$ and $\eta_r$ are the efficiencies of the transmit and receive antennas. 
The parameter $L_{\extra}$ includes possible extra losses.
Letting $n_{\mathrm{s},\mathrm{margin}}$ be the link margin, the difference $n^*_\mathrm{s}\,[\mathrm{dB phe/ns}] = n_\mathrm{s}\,[\mathrm{dB phe/ns}] - n_{\mathrm{s},\mathrm{margin}}\,[\mathrm{dB phe/ns}]$ represents the value of the number of received signal photons for which, for given average number of background noise photons $n_\mathrm{b}$, the \ac{SCPPM} scheme must guarantee the target performance over the considered channel, e.g., Poisson \ac{PPM} if photon counting detection is used \cite{gagliardi1976:optical}.
A suitable combination of the \ac{SCPPM} parameters $M$, $T_\mathrm{s}$, and code rate $R$ (see Appendix~\ref{appendix:SCPPM}) can then be found to fulfill the constraints, yielding the achieved data rate $B_r$.
Denoting by $N_1$ the number of CRC and termination bits and by $T_F$ the time duration of the coded \ac{PPM} channel block, we have 
\begingroup
\allowdisplaybreaks
\begin{align}
B_r  &=  \frac{15120\, R-N_1}{T_F} = \frac{15120\, R-N_1}{M\,T_\mathrm{s}\,(15120/\log_2M) \, \alpha_{\mathrm{GT}}}
\end{align}
\endgroup
where the factor $\alpha_{\mathrm{GT}}>1$ captures the guard times. For instance, with a $25\%$ guard time (with respect to the coded PPM block time), we have $\alpha_{\mathrm{GT}}=1.25$.

Estimation of the background noise is performed by computing the noise photons per second, $n_\mathrm{b}$, through
\begingroup
\allowdisplaybreaks
\begin{align}
n_\mathrm{b}  &=  \frac{ P_{\mathrm{source}}} {h\,f}
\end{align}
\endgroup
where $P_{\mathrm{source}}$ is the optical noise power generated by the considered noise source (sun, planet, star), $h$ the Plank’s constant, and $f$ the radiation frequency. 
Values of $P_{\mathrm{source}}$ have been estimated using the \ac{ITU-R} approach provided in \cite{ITU:06}. 
The Sun direct noise is found to be several orders of magnitude greater than that from other sources. 
For this reason, we assume that no optical communication can be established between the two \acp{S/C} when any portion of the Sun is in the \ac{FOV} of the receiver. 
When the instrument is free of the Sun direct noise, the only considered noise contribution is the solar radiation reflected from the central body of the deep space probe orbit as a function of its albedo. This, in turn, is computed assuming the planet fully illuminated and entirely contained within the instrument \ac{FOV} (worst-case).

\section{Results}\label{sec:results}

This section is devoted to the analysis of the performance of two-leg deep space relay systems, along with comparison with systems based on a direct X-band link (Section~\ref{subsec:direct}). In particular, Section~\ref{subsec:transparent_results} addresses the case of a transparent relay, while the case of a regenerative relay is considered in Section~\ref{subsec:regenerative_results}, including both \ac{EHF} and optical bands in leg-1.

\begin{figure*}[t]
\centering
\subfigure{
    \includegraphics[width=0.7\columnwidth]{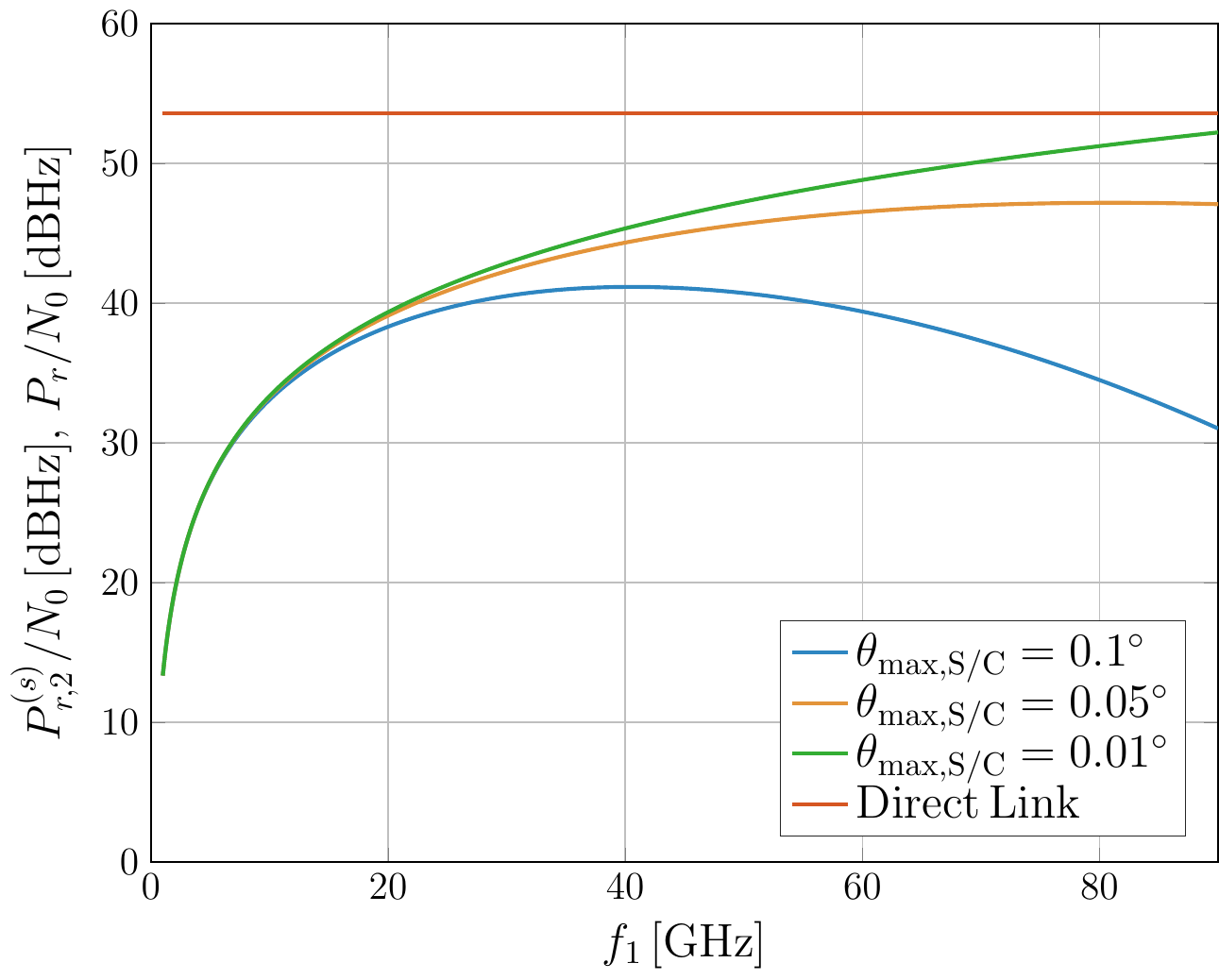}
}
\hspace{8mm}
\subfigure{
    \includegraphics[width=0.7\columnwidth]{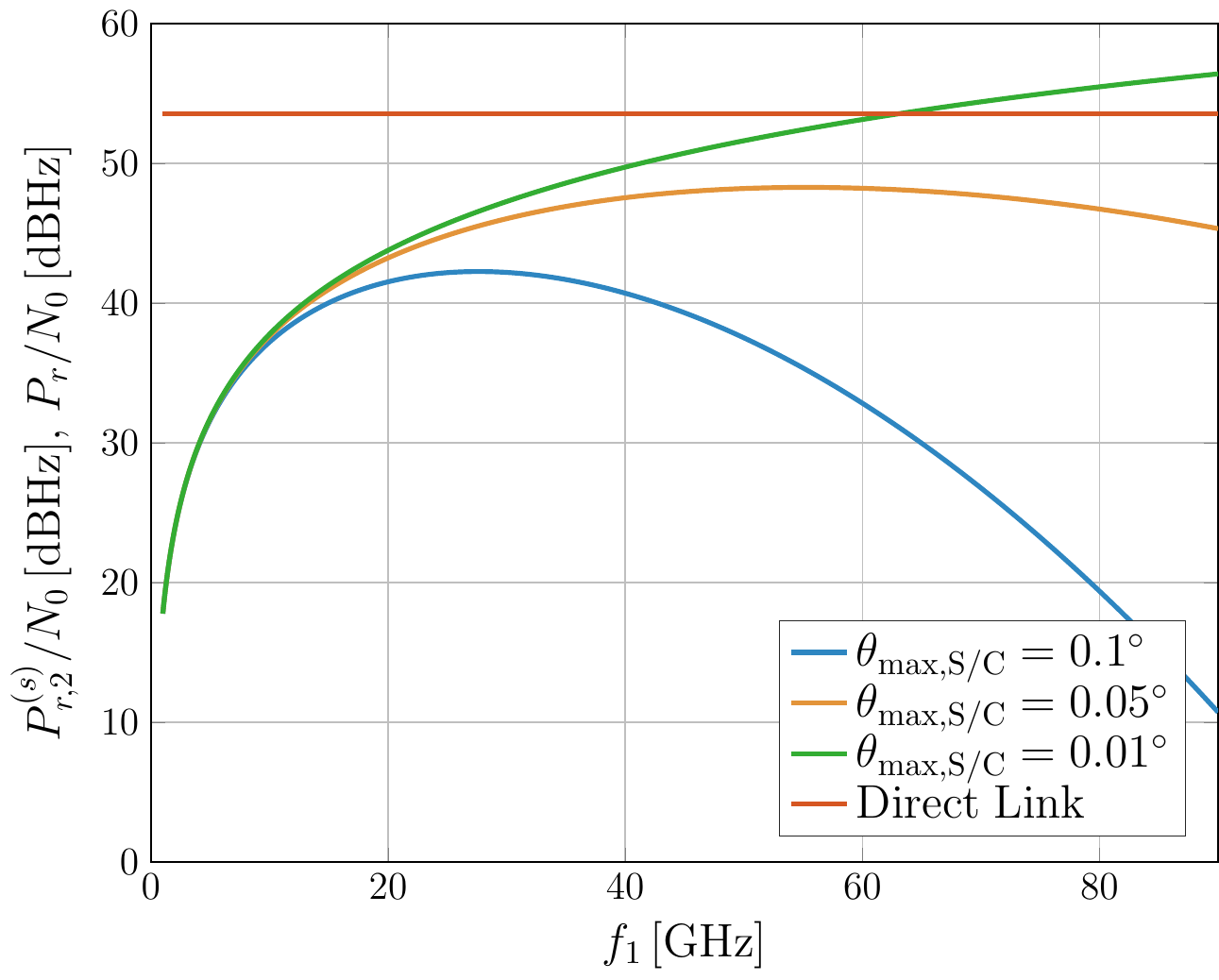}
}
    \caption{$P_{r,2}^{(s)}/N_0$ values of a two-leg system with transparent relay versus the leg-1 frequency $f_1$, compared with the $P_r/N_0$ value of a direct link X-band system (red horizontal line); $95\%$ availability, relay receive antenna diameter $d_{r,1}=3\,\mathrm{m}$ (left) and $d_{r,1}=5\,\mathrm{m}$ (right), several pointing error values. Mars scenario.}
    \label{fig:Transparent_Mars}
\end{figure*}

\subsection{Results for Transparent Relay}\label{subsec:transparent_results}

In the transparent relay case we use the framework discussed in Section~\ref{subsec:transparent_framework}, with the goal of investigating the range of values of the leg-1 frequency $f_1$ for which the two-leg configuration can outperform the direct one, for the same input values (the ones summarized in Table~\ref{table:Venus_table}, including the transmit power of the deep space probe), the same \ac{G/S} ($35\,\mathrm{m}$ \ac{DSA}) and for a leg-2 frequency $f_2=27\,\mathrm{GHz}$. 
The value $f_2=27\,\mathrm{GHz}$ belongs to the recommended range for near-Earth space-to-Earth communication \cite{ECSS2011:50-05C}. A GEO relay is considered. 

The pointing losses have been estimated through the method based on $\theta_{\max}$. 
The values of the system noise temperature in the relay and in the G/S have been set to $T_{\syst,1}=T_{\syst,2}=100\,\mathrm{K}$.  Finally, the atmospheric attenuation at $27\,\mathrm{GHz}$, assuming the \ac{DSA}-1 location (New Norcia), has been computed using the \ac{ITU-R} recommendations; the obtained value is $1.33\,\mathrm{dB}$ for $95\%$ availability with an elevation angle $\theta=53.89$°, corresponding to a GEO relay. 
The results have been obtained for an \ac{EIRP} of the GEO relay equal to $62\,\mathrm{dBW}$.  
It should be remarked, however, that the results exhibit a very low sensitivity to the relay \ac{EIRP}, since the two-leg system performance is limited essentially by the first link. 

Fig.~\ref{fig:Transparent_Mars} shows the \ac{G/S} $P_r/N_0$ value of the direct X-band link and the \ac{G/S} $P_{r,2}^{(s)}/N_0$ values of the two-leg system, given by the right hand side of \eqref{eq:transparent_SN0}, for a link availability of $95\%$ and for a relay receive antenna diameter $d_{r,1}$ of $3\,\mathrm{m}$ and $5\,\mathrm{m}$, respectively. 
The two parameters are plotted versus the leg-1 frequency $f_1$ (hence the direct link $P_r/N_0$ remains constant). 
In both figures, curves for the two-leg system are shown for values of the maximum pointing error $\theta_{\max}=0.1$°, $\theta_{\max}=0.05$°, and $\theta_{\max}=0.01$° in the \acp{S/C} (deep space probe and relay, both in transmission and reception), while $\theta_{\max} = 10^{-4}\,\mathrm{rad} \approx 0.0057$° has been assumed in all cases at the \ac{G/S}. 
The considered scenario is Mars, but very similar results are obtained in the other cases. 

The $P_r/N_0$ of the direct link remains essentially the same for all maximum pointing error values, owing to the use of the X-band. As it is possible to observe, when the relay antenna diameter equals $3\,\mathrm{m}$, the $P_{r,2}^{(s)}/N_0$ values for the two-leg system are lower than the ones of the direct X-band link for all values of $f_1$ in the range of interest, and regardless of the pointing errors.
For a relay receive antenna diameter of $5\,\mathrm{m}$ and only for a spacecraft maximum pointing error of $\theta_{\max}=0.01$°, the two-leg system exhibits some gain over the direct link one. 
More in detail, for $f_1 \approx 63\,\mathrm{GHz}$, the $P_{r,2}^{(s)}/N_0$ values of the two-leg system matches the $P_r/N_0$ value of the direct link system.
In this situation, if the two systems use the same coding and modulation, they achieve the same data rate $B_r$ with the same reliability.
For frequencies $f_1$ larger than this value, the two-leg system exhibits a gain that may be used to decrease the ground antenna size for the same data rate.
We observe, however, that the gain is marginal so that the transparent configuration seems not a viable solution.
More encouraging results have been found for regenerative relays, as addressed in the following. 

\begin{figure*}[t]
\centering
\subfigure{
    \includegraphics[width=0.7\columnwidth]{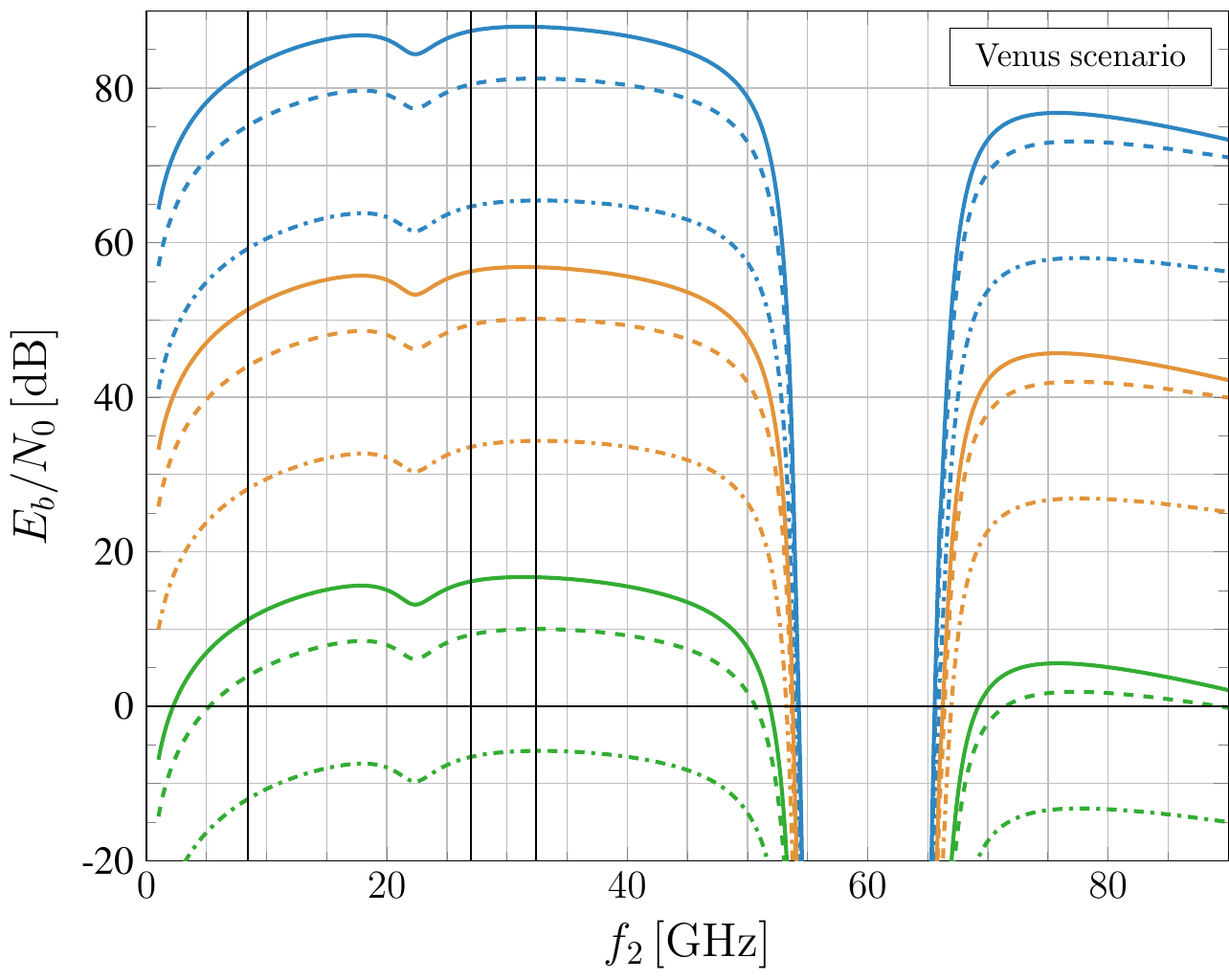}
}
\hspace{8mm}
\subfigure{
    \includegraphics[width=0.7\columnwidth]{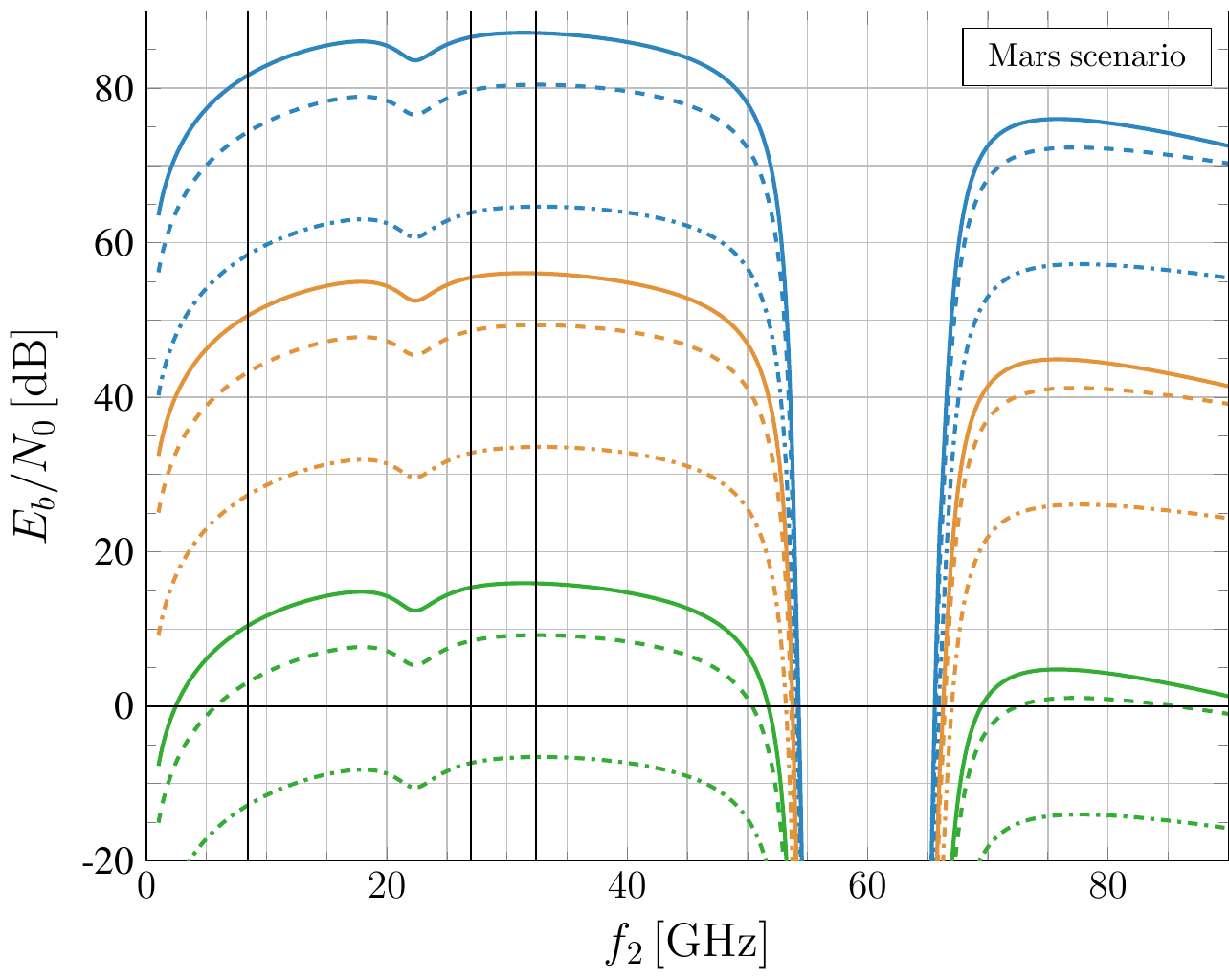}
}
\subfigure{
    \includegraphics[width=0.7\columnwidth]{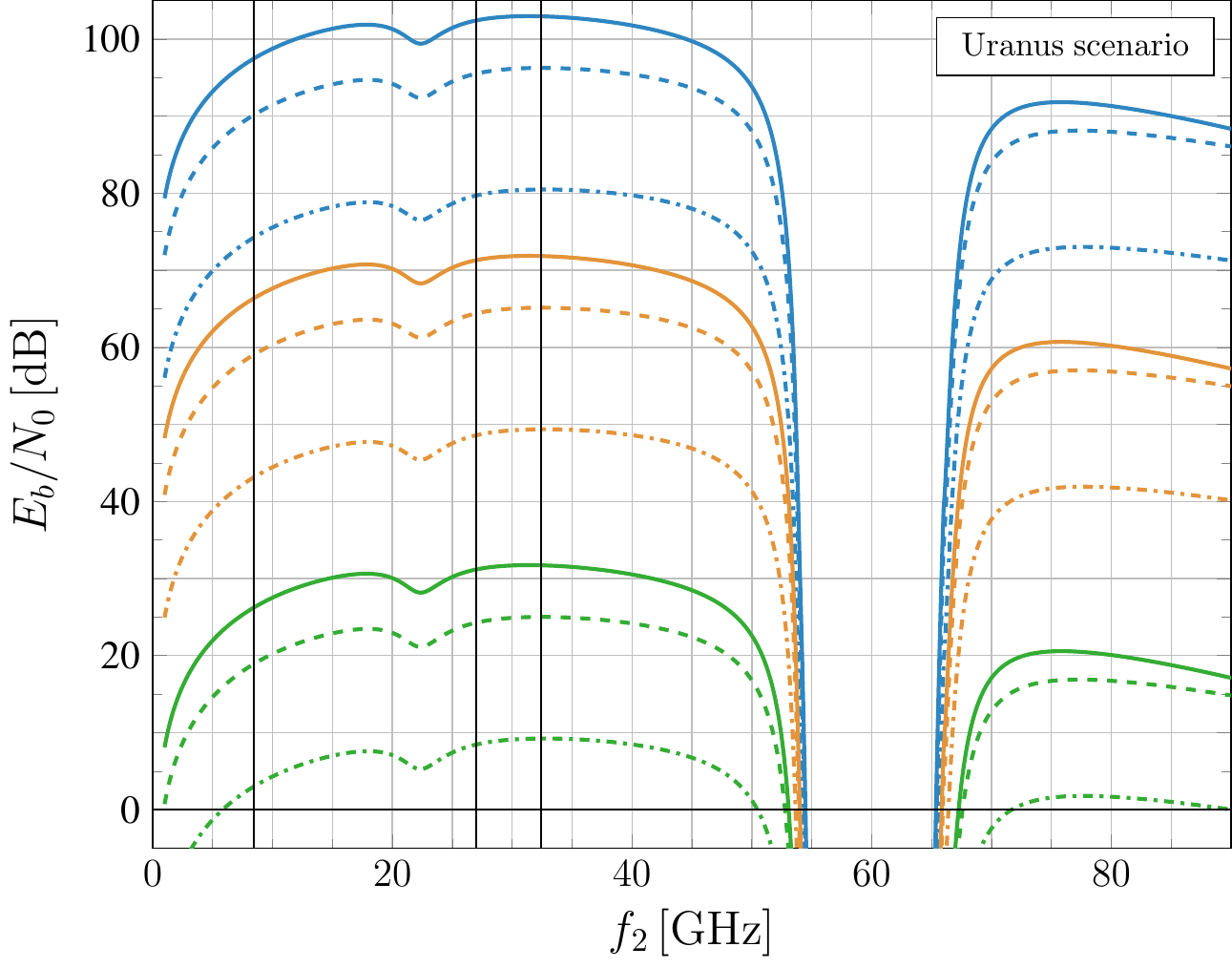}
}
\hspace{8mm}
\subfigure{
    \includegraphics[width=0.7\columnwidth]{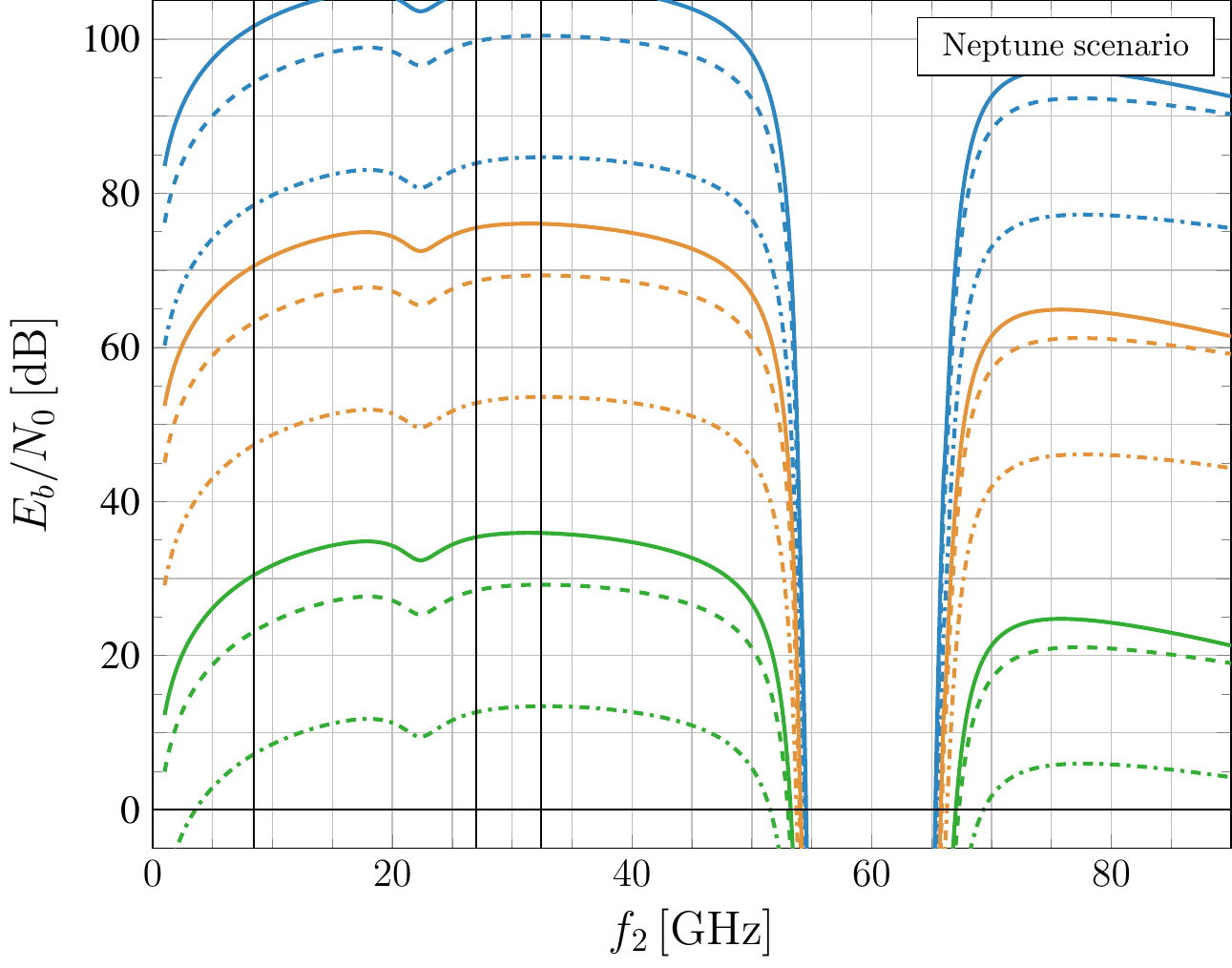}
}
    \caption{Available $E_b/N_0$ values at the \ac{G/S} for leg-2 in a two-leg system with regenerative relay, versus the leg-2 frequency $f_2$. Data rates: $100\,\mathrm{kbps}$ (Venus), $120\,\mathrm{kbps}$ (Mars), $3.15\,\mathrm{kbps}$ (Uranus), $1.2\,\mathrm{kbps}$ (Neptune). Blue: GEO relay; orange: L1 relay; green: L4/L5 relay. Solid: G/S antenna diameter $35\,\mathrm{m}$; dashed: G/S antenna diameter $15\,\mathrm{m}$; dot-dashed: G/S antenna diameter $2.4\,\mathrm{m}$. The black vertical lines correspond to frequencies $8.42\,\mathrm{GHz}$, $27\,\mathrm{GHz}$, and $32.4\,\mathrm{GHz}$.}
    \label{fig:Leg2Br}
\end{figure*}

\subsection{Results for Regenerative Relay}\label{subsec:regenerative_results}

As pointed out in Section~\ref{subsec:PeTot}, in the regenerative case we can analyze the two links separately.
We start with the analysis of leg-2, showing that this link is not critical for the considered data rate.
We then focus on leg-1, including both the \ac{EHF} and the optical case.

\subsubsection{Leg-2 Analysis}\label{subsubsec:leg2}
We provide an analysis of leg-2 (relay-to-ground link) considering three different relay options, namely, a GEO relay, a relay placed in Lagrange point L1, and a relay placed in Lagrange points L4/L5. 
We assume a relay transmit power $P_{t,2}=65\,\mathrm{W}$, transmit antenna diameter $d_{t,2}=2\,\mathrm{m}$, efficiency $\eta_{t,2}=0.7$, and transmit implementation loss $A_{t,2}=1.5\,\mathrm{dB}$, corresponding to an \ac{EIRP} of approximately $70\,\mathrm{dBW}$ at $27\,\mathrm{GHz}$. 
The receiver system noise temperature is set to $T_{\syst} = 100\,\mathrm{K}$. 
Atmospheric losses are computed for a percentage of link availability equal to $95\%$ and 10° elevation in all cases. 
Pointing losses are computed using the deterministic model, assuming maximum pointing errors $\theta_{\max,\text{G/S}} = 100 \text{ μrad} \approx 0.0057$° at the \ac{G/S} and $\theta_{\max} = 0.05°$ at the relay.  

Fig.~\ref{fig:Leg2Br} illustrates the available $E_b/N_0$ in each of the scenarios of interest. With suppressed carrier modulation, the available $E_b/N_0$ can be computed as the ratio of $P_r/N_0$ (the same for all scenarios) to the data rate $B_r$. To this aim, we use the data rate values shown in Table~\ref{table:data_rates} (X-band direct link data rates) and summarized hereafter: $100\,\mathrm{kbps}$ (Venus), $120\,\mathrm{kbps}$ (Mars), $3.15\,\mathrm{kbps}$ (Uranus), $1.2\,\mathrm{kbps}$ (Neptune). 
Blue curves in Fig.~\ref{fig:Leg2Br} are relevant to a GEO relay, orange curves to an L1 relay, and green curves to an L4/L5 relay. 
Moreover, solid, dashed, and dot-dashed curves correspond to a \ac{G/S} antenna diameter of $35\,\mathrm{m}$, $15\,\mathrm{m}$, and $2.4\,\mathrm{m}$, respectively. 

\begin{figure*}[t]
\centering
\subfigure{
    \includegraphics[width=0.7\columnwidth]{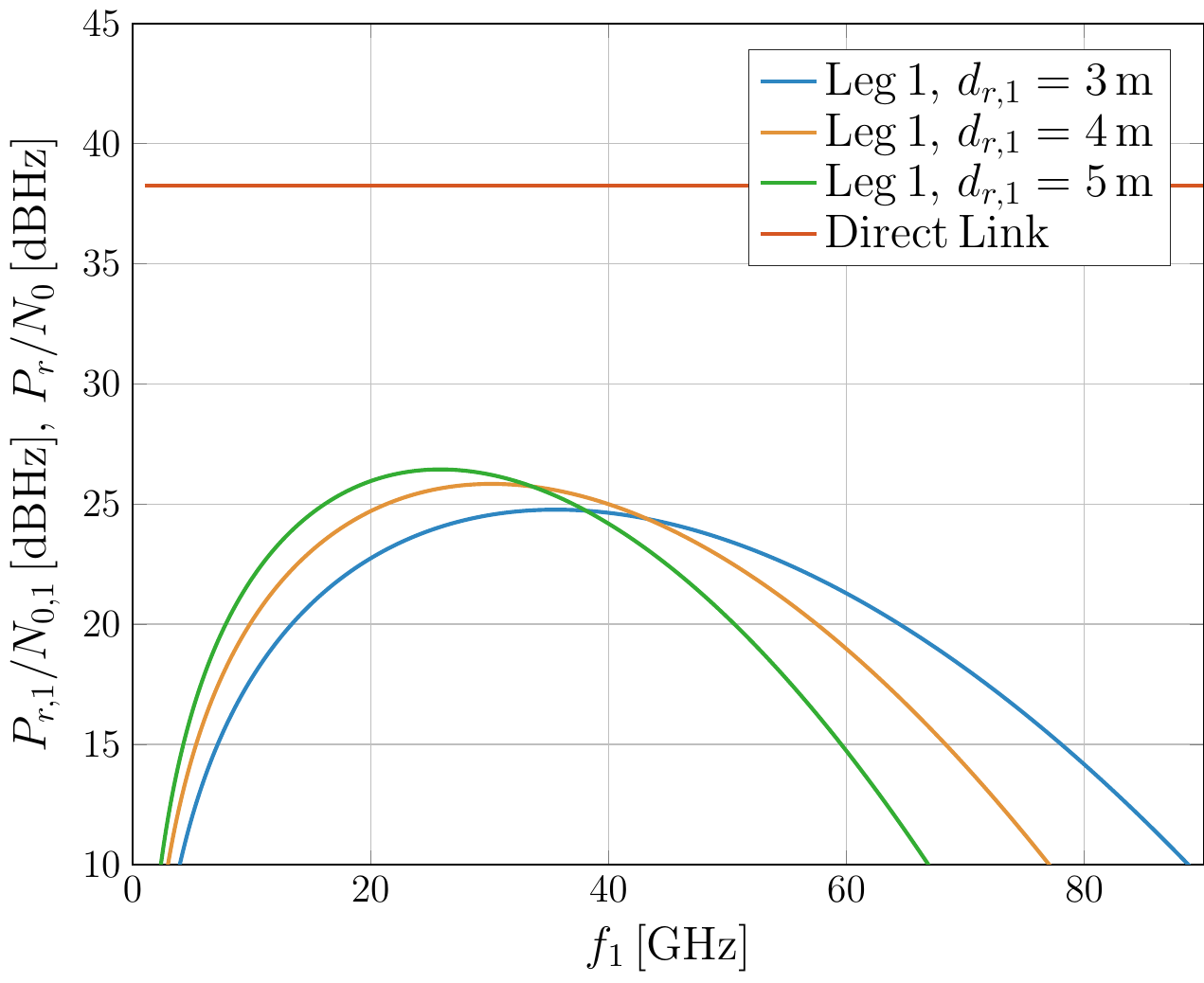}
}
\hspace{8mm}
\subfigure{
    \includegraphics[width=0.7\columnwidth]{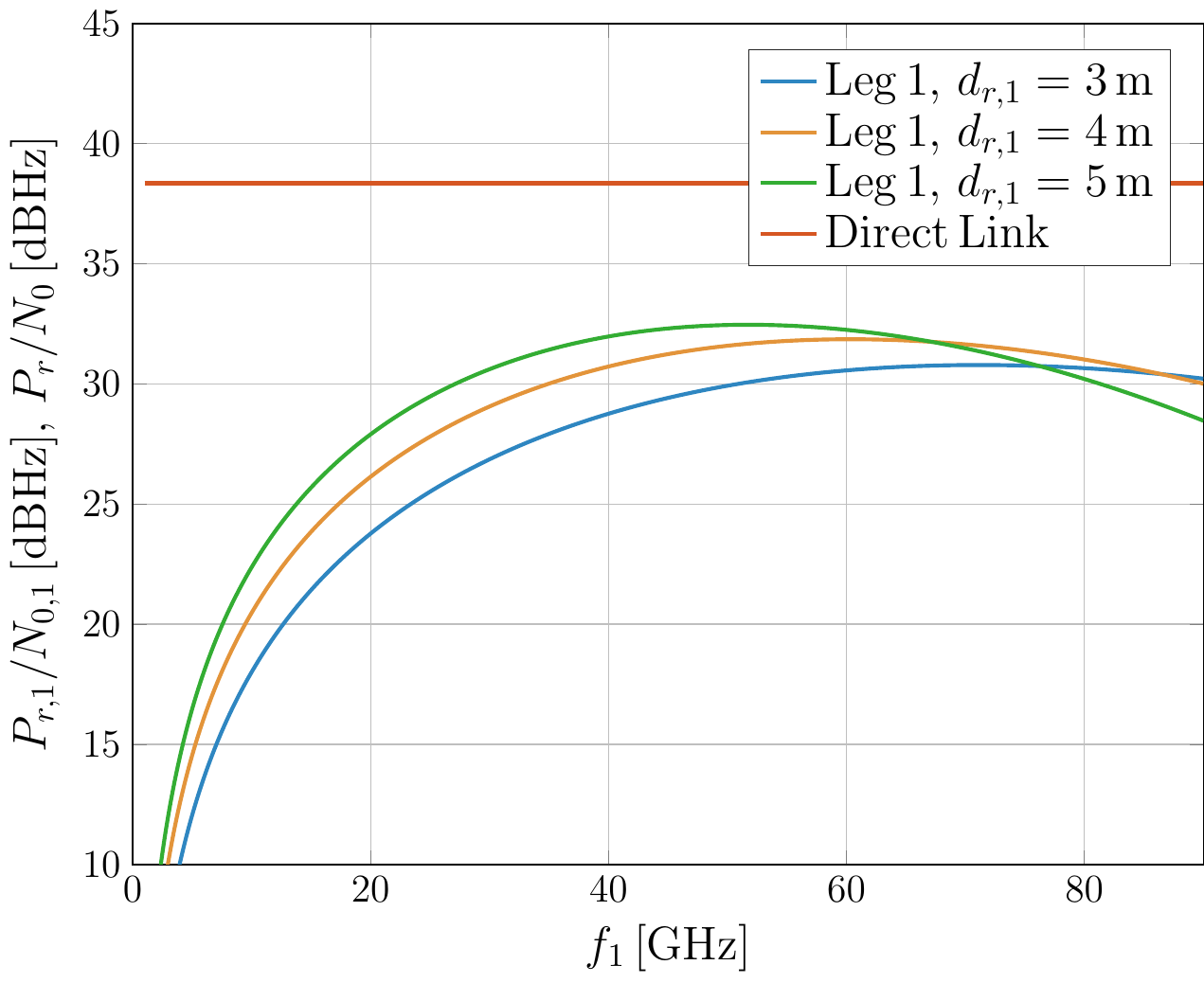}
}
    \caption{$P_{r,1}/N_{0,1}$ values of a two-leg system with regenerative relay versus the leg-1 frequency $f_1$, compared with the $P_r/N_0$ values of direct link X-band system (red horizontal line); several relay receive antenna diameters; S/C (deep space and relay) pointing accuracy $\theta_{\max}=0.1$° (left) and $\theta_{\max}=0.05$° (right); \ac{G/S} pointing accuracy $\theta_{\max,\mathrm{G/S}}=10^{-4}\,\mathrm{rad}$. Uranus scenario.}
    \label{fig:Leg1Uranus}
\end{figure*}

Assuming $(E_b/N_0)^* = 4\,\mathrm{dB}$, corresponding to a concatenated scheme employing a punctured $(7,3/4)$ convolutional code and a $(255,223)$ RS code with interleaving depth $16$, and an $E_b/N_0$ margin of $4\,\mathrm{dB}$, the available $E_b/N_0$ at the \ac{G/S} should be at least $8\,\mathrm{dB}$. 
As we can observe looking at Fig.~\ref{fig:Leg2Br}, this is not a concern for a GEO or a Lagrange L1 relay, regardless of the G/S antenna size, in the whole $f_2$ frequency range, apart from frequencies around $60\,\mathrm{GHz}$.
The large observed margins are due to the low data rates since, as explained previously, we are targeting the same data rate in both legs.

As Lagrange L4/L5 points are in deep space, in case of a Lagrange L4/L5 relay a frequency $f_2$ in X-band or Ka-band should be considered. 
Using as channel coding scheme the rate-$1/4$ turbo code with information block length $7136$ bits, the minimum $E_b/N_0$ to achieve a target \ac{BER} of $10^{-6}$ is $(E_b/N_0)^*=0.23\,\mathrm{dB}$, so we need at least $4.23$ dB to achieve an $E_b/N_0$ margin of $4\,\mathrm{dB}$. 
As we can see from Fig.~\ref{fig:Leg2Br}, in the Uranus and Neptune scenarios use of the Ka-band allows meeting the $4.23\,\mathrm{dB}$ requirement even with a \ac{G/S} antenna diameter of $2.4\,\mathrm{m}$. 
More in detail, the available $E_b/N_0$ is equal to $9.25\,\mathrm{dB}$ in the Uranus scenario and to $13.43\,\mathrm{dB}$ in the Neptune one. 
This is instead not the case in the Venus and Mars scenarios, where the higher data rate makes the link margin poorer: Considering again the Ka-band and a G/S antenna diameter of $2.4\,\mathrm{m}$, we have $E_b/N_0=-5.77\,\mathrm{dB}$ for Venus and $E_b/N_0=-6.56\,\mathrm{dB}$ for Mars. In the Venus and Mars scenarios, we can calculate the minimum G/S antenna size to fulfil the $E_b/N_0 \geq 4.23$ dB requirement assuming $f_2=32.4$ GHz, $95\%$ availability and $10$° elevation. We obtain a \ac{G/S} antenna diameter of $7.7\,\mathrm{m}$ for the Venus scenario and of $8.4\,\mathrm{m}$ for the Mars scenario. 
These values are remarkably lower that the DSA ones ($35\,\mathrm{m}$), and are upper bounds to the case of $90\%$ availability and $20$° elevation often targeted for the Ka-band. 
Repeating the analysis for the X-band ($f_2=8.42\,\mathrm{GHz}$), considering again $95\%$ availability and $10$° elevation we obtain $15.7\,\mathrm{m}$ for Venus and $17.2\,\mathrm{m}$ for Mars.

\subsubsection{\ac{EHF} Leg-1 Analysis}
The analysis of leg-1 is carried out by comparing the \ac{G/S} $P_r/N_0$ values of a direct X-band link with the $P_{r,1} \/N_{0,1}$ values characterizing leg-1 in the two-leg system. 
A GEO relay is considered, although the obtained results remain essentially valid for the other relay options since the leg-1 range has very little sensitivity to the relay position, even in Lagrange points. 
The input values are the same for both systems (direct link and two-leg) and are the ones summarized in Table~\ref{table:Venus_table}. 
The system noise temperature is assumed equal to $T_{\syst}=100\,\mathrm{K}$ both in the ground receiver and in the relay receiver. 
Coherently with the direct link analysis (Section~\ref{subsec:direct}), the implementation losses are equal to $1.5\,\mathrm{dB}$ in space (deep space probe transmitter, relay receiver, relay transmitter) and $0.5\,\mathrm{dB}$ on ground. 
The X-band atmospheric attenuation impairing the direct link, computed for $95\%$ availability, $10$° elevation, and the DSA-1 location, is $A_{\atm}=0.5\,\mathrm{dB}$. 
For both the direct X-band link and the interplanetary leg-1 of the two-leg system, the pointing losses have been estimated using the approach based on $\theta_{\max}$; the values $\theta_{\max}=0.1$°, $\theta_{\max}=0.05$°, and $\theta_{\max}=0.01$° of maximum pointing error have been assumed for \acp{S/C} (deep space probe in transmission and relay in reception), while $\theta_{\max,\mathrm{G/S}}=100$ μrad $\approx 0.0057$° has been assumed at the \ac{G/S}.

The goal is to investigate the value of the frequency $f_1$ for which the $P_{r,1} / N_{0,1}$ value of the S/C-to-relay link is equal to the $P_r/N_0$ of the direct X-band link. 
When this condition is met, using the same modulation and coding techniques, the direct link and leg-1 achieve the same data rate with the same reliability.  
As addressed above, leg-2 is not a concern; therefore, the data rate is achieved by the whole two-leg system, with \ac{G/S} antennas smaller than DSA ones.

\begin{figure*}[t]
\centering
\subfigure{
    \includegraphics[width=0.7\columnwidth]{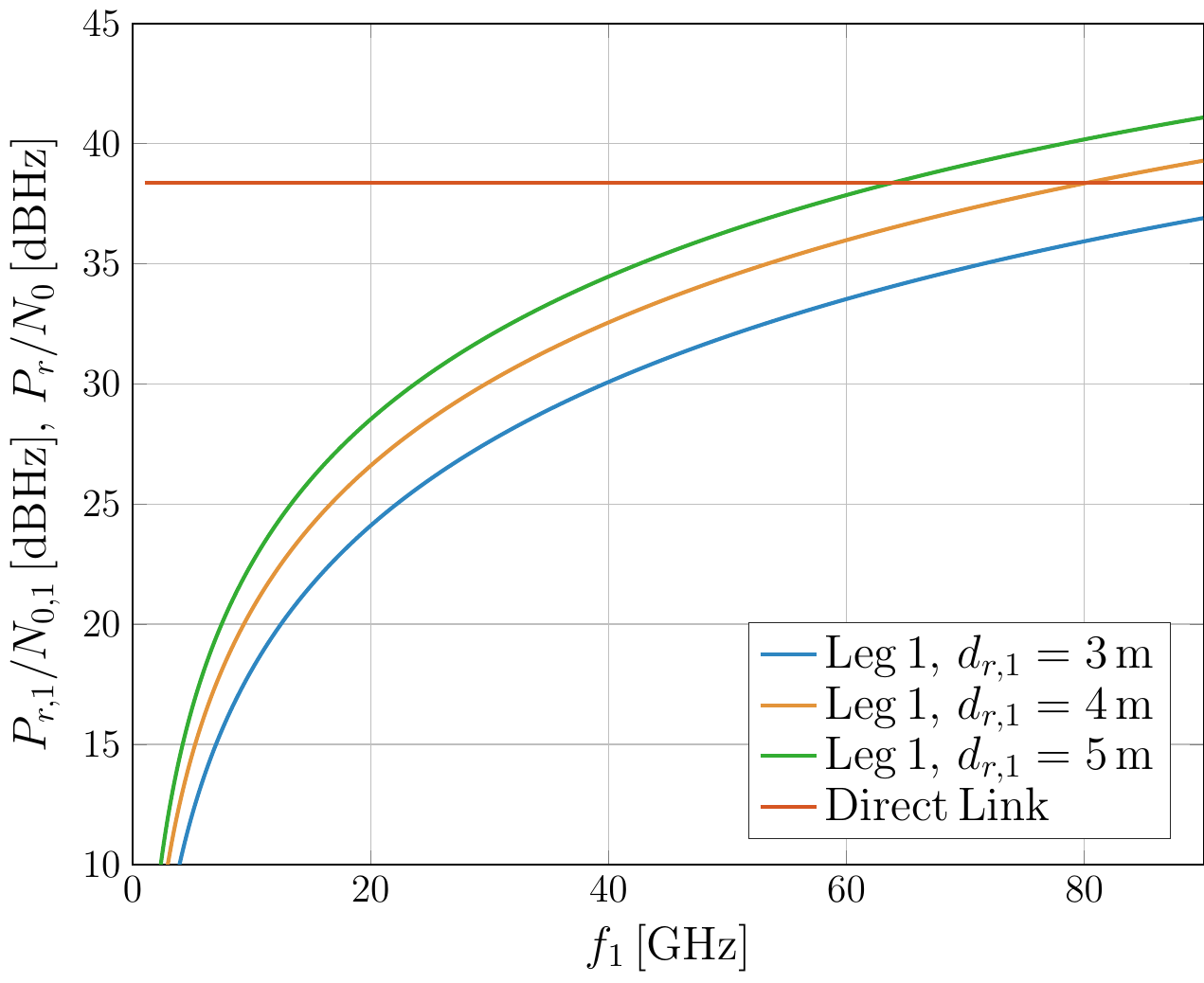}
}
\hspace{8mm}
\subfigure{
    \includegraphics[width=0.7\columnwidth]{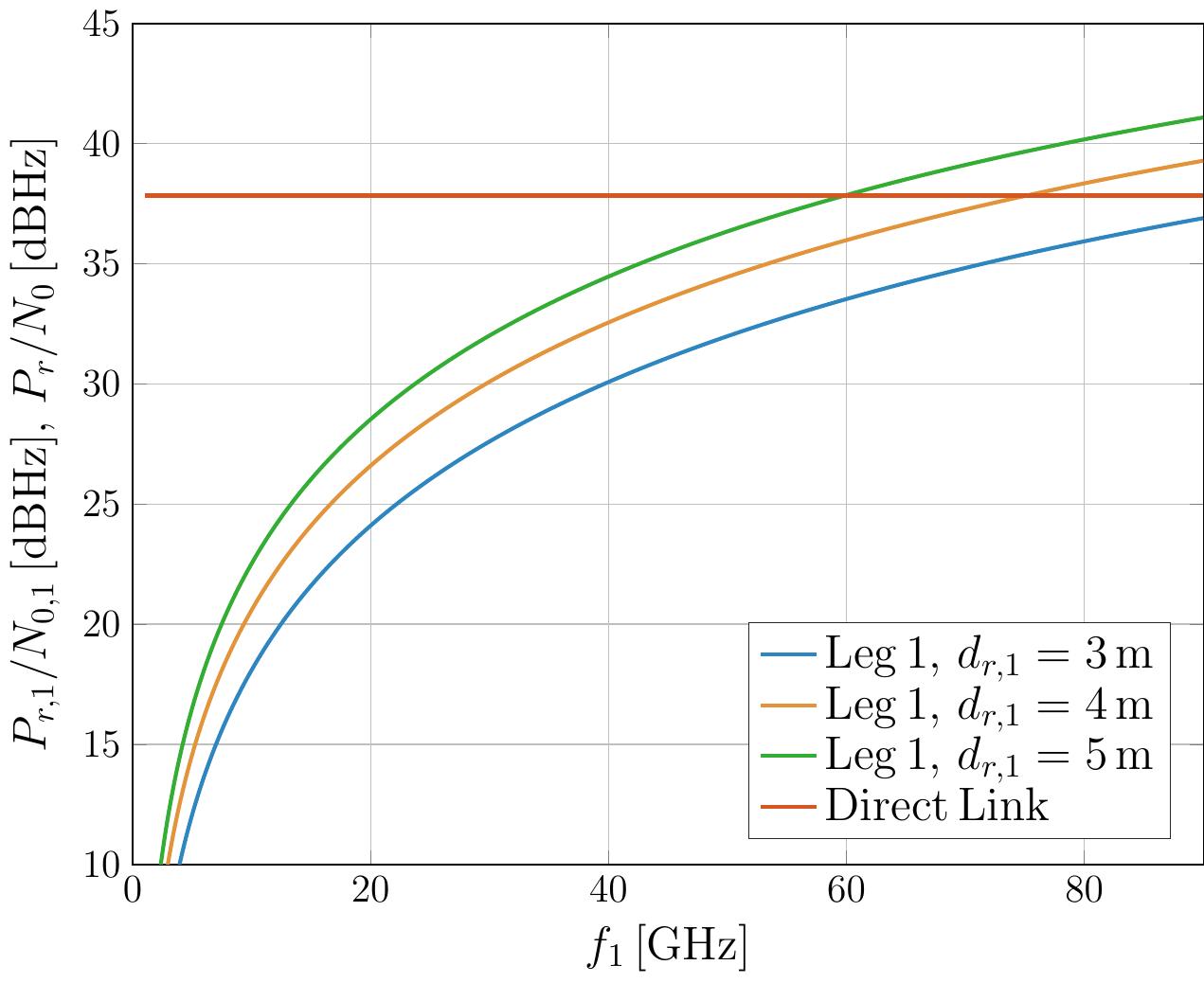}
}
\subfigure{
    \includegraphics[width=0.7\columnwidth]{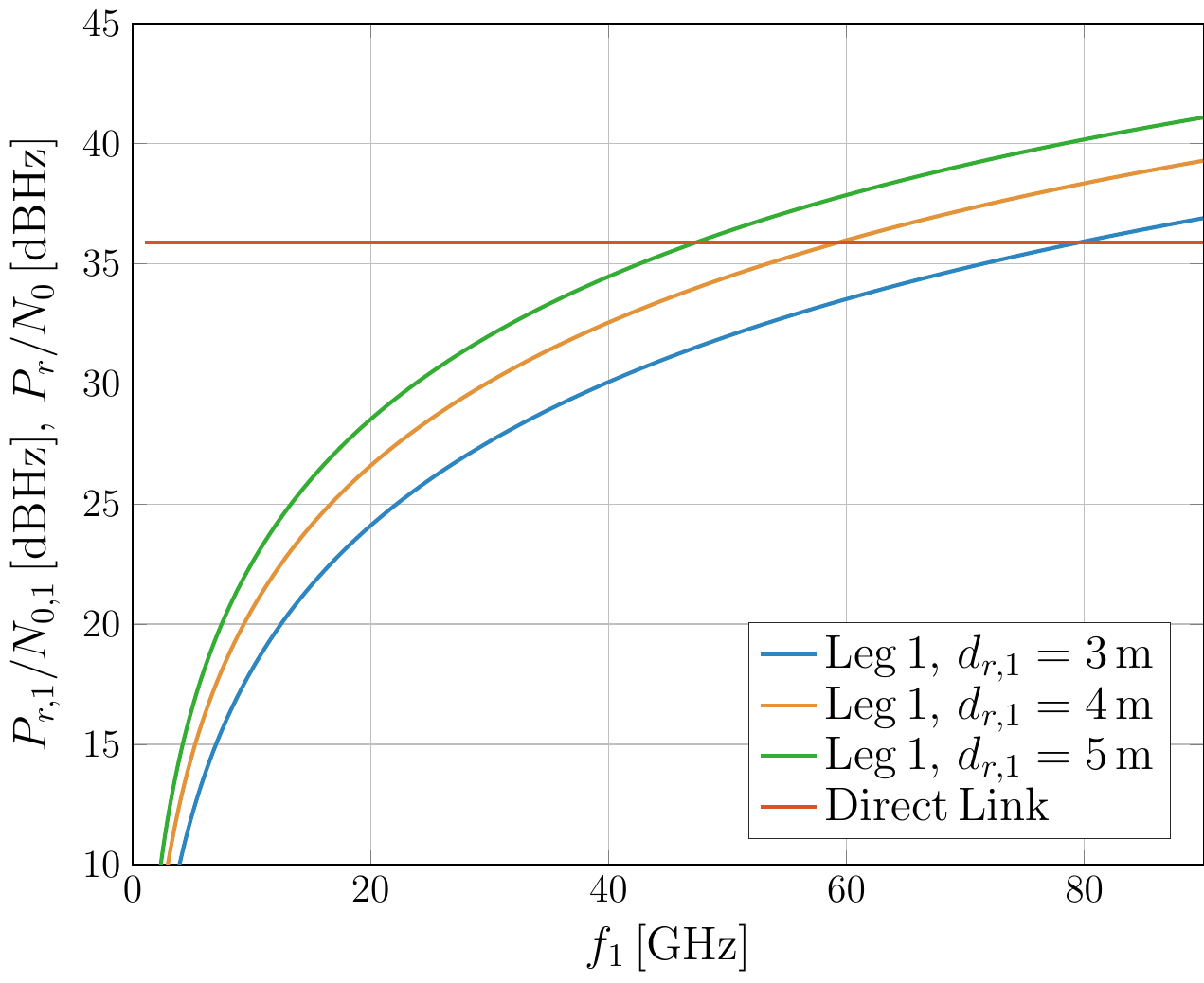}
}
\hspace{8mm}
\subfigure{
    \includegraphics[width=0.7\columnwidth]{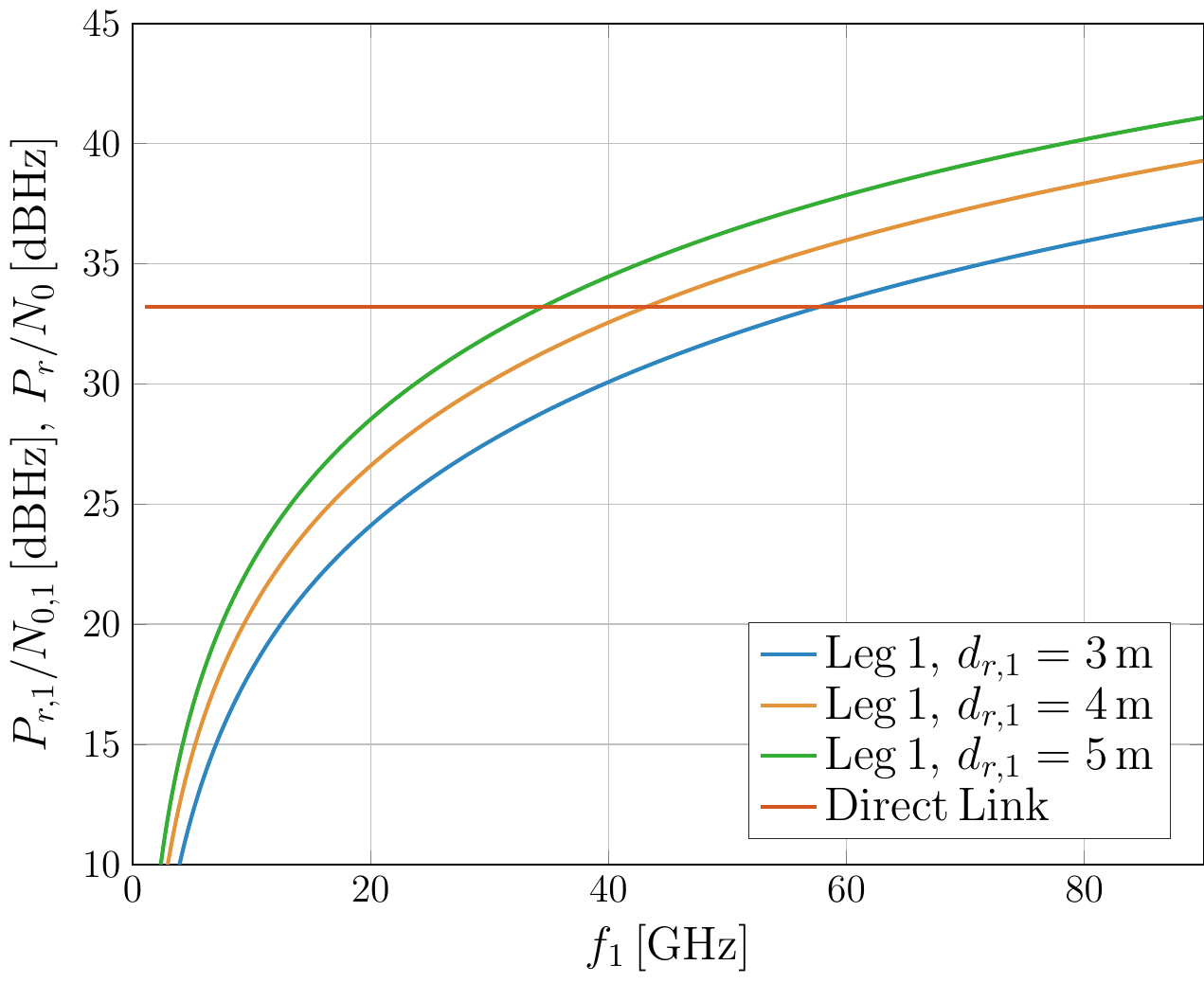}
}
    \caption{$P_{r,1}/N_{0,1}$ values of a two-leg system with regenerative relay versus the frequency $f_1$ in leg 1, compared with the $P_r/N_0$ values of direct link X-band system (red horizontal line); several relay receive antenna diameters; S/C (deep space and relay) pointing accuracy $\theta_{\max}=0.01$° and \ac{G/S} pointing accuracy $\theta_{\max,\mathrm{G/S}}=10^{-4}\,\mathrm{rad}$. Top-left: $95\%$ availability; Top-right: $99\%$ availability; Bottom-left: $99.9\%$ availability; Bottom-right: $99.99\%$ availability. Uranus scenario.}
    \label{fig:Leg1Uranus_availability}
\end{figure*}

In Fig.~\ref{fig:Leg1Uranus} and Fig.~\ref{fig:Leg1Uranus_availability}, the $P_r/N_0$ values of the direct X-band link are compared with the $P_{r,1} / N_{0,1}$ values of leg-1 in a two-leg system, for $f_1$ ranging from $1\,\mathrm{GHz}$ to $90\,{\mathrm{GHz}}$ and for three different values of the relay receiving antenna diameter $d_{r,1}$, namely, $3\,\mathrm{m}$, $4\,\mathrm{m}$, and $5\,\mathrm{m}$. In particular, Fig.~\ref{fig:Leg1Uranus} assumes a maximum pointing error $\theta_{\max} = 0.1$° and $\theta_{\max} = 0.01$° in the spacecrafts, while Fig.~\ref{fig:Leg1Uranus_availability} assumes $\theta_{\max}=0.01$°. 
This allows emphasizing the effect of spacecraft pointing errors on the system performance. 
The considered scenario is Uranus, although results for the other scenarios are very similar and lead essentially to the same conclusions.

\begin{table}[!t]
\begin{center}
\caption{Values of crossing frequency $f_1$ for different availabilities and different diameters of the relay receiving antenna.}\label{table:Crossing}
\footnotesize{
\begin{tabular}{ccc}
\toprule
 {\textbf{Relay antenna diameter}} & {\textbf{Availability}} & {\textbf{Crossing frequency $f_1$}}\\
 \toprule
 \multirow{6}{*}{$3.0\,\mathrm{m}$} & $95.00\%$ & $108.3\,\mathrm{GHz}$\\
    & $97.00\%$ & $105.6\,\mathrm{GHz}$\\
	& $99.00\%$ & $101.2\,\mathrm{GHz}$\\
 	& $99.50\%$	& $95.4\,\mathrm{GHz}$\\
	& $99.90\%$	& $79.6\,\mathrm{GHz}$\\
	& 99.99\%	& $57.7\,\mathrm{GHz}$\\
	\midrule
 \multirow{6}{*}{$4.0\,\mathrm{m}$} & $95.00\%$	& $80.3\,\mathrm{GHz}$\\
	& $97.00\%$	& $78.4\,\mathrm{GHz}$\\
	& $99.00\%$	& $75.2\,\mathrm{GHz}$\\
	& $99.50\%$	& $71.0\,\mathrm{GHz}$\\
	& $99.90\%$	& $59.3\,\mathrm{GHz}$\\
	& $99.99\%$	& $43.2\,\mathrm{GHz}$\\
	\midrule
 \multirow{6}{*}{$5.0\,\mathrm{m}$} & $95.00\%$	& $63.9\,\mathrm{GHz}$\\
	& $97.00\%$	& $62.4\,\mathrm{GHz}$\\
	& $99.00\%$	& $59.9\,\mathrm{GHz}$\\
	& $99.50\%$	& $56.6\,\mathrm{GHz}$\\
	& $99.90\%$	& $47.4\,\mathrm{GHz}$\\
	& $99.99\%$	& $34.5\,\mathrm{GHz}$\\
\bottomrule
\end{tabular}}
\end{center}
\end{table}

As it can be immediately observed from Fig.~\ref{fig:Leg1Uranus}, the two-leg system is outperformed by the X-band direct link one for spacecraft maximum pointing errors $\theta_{\max}=0.1$° and $\theta_{\max}=0.05$°. 
This is because the $P_{r,1} / N_{0,1}$ value at the relay receiver is upper bounded by the direct link $P_r/N_0$ at the \ac{G/S} over the whole range of frequencies $f_1$. 
The situation is different when a better pointing accuracy is met between the deep space probe and the relay. 
As shown in Fig.~\ref{fig:Leg1Uranus_availability}, in fact, when $\theta_{\max}=0.01$° there exist frequencies $f_1$, whose values depend on the relay receive antenna diameter, above which leg-1 of the two-leg system outperforms the X-band direct link in terms of $P_{r,1} / N_{0,1}$ compared with the \ac{G/S} $P_r/N_0$ of the direct link (hence, in terms of data rate for the same coding, modulation, and reliability requirement). 
Fig.~\ref{fig:Leg1Uranus_availability} also elaborates on the result by varying the percentage of direct X-band link availability, considering $99\%$, $99.9\%$, and $99.99\%$ besides the standard value $95\%$, the maximum pointing error at the \ac{G/S} being $\theta_{\max,\mathrm{G/S}}=100$~μrad in all cases. 
As it can be seen, the effect of a higher availability is to reduce the crossover frequency $f_1$ beyond which leg-1 outperforms the direct link. 
The values of crossover frequencies $f_1$ corresponding to the four charts in Fig.~\ref{fig:Leg1Uranus_availability} are summarized in Table~\ref{table:Crossing}, in which also the availability values $97\%$ and $99.5\%$ have been considered for the sake of completeness. 
As we can see, a relay receive antenna diameter of $3\,\mathrm{m}$ can achieve a gain over the direct link system, only for frequencies $f_1$ that are larger that about $80\,\mathrm{GHz}$, unless an unusually high availability of $99.99\%$ is targeted. 
In contrast, a relay receive antenna diameter of $4\,\mathrm{m}$ with an $f_1$ frequency in the range $71$-$80\,\mathrm{GHz}$ or a relay receive antenna diameter of $5\,\mathrm{m}$ with an $f_1$ frequency in the range $56$-$64\,\mathrm{GHz}$ are the minimum requirements for the two-leg system to achieve a gain over the X-band direct one, assuming values of availability ranging between $95\%$ and $99.5\%$. 

\subsubsection{Optical Leg-1 Analysis}\label{subsubsec:optical_results}

We now analyze the case of an optical leg-1. 
We start by analyzing the maximum range that can be achieved for different values of the pointing accuracy $\sigma_\theta$ (or $\theta_\mathrm{max}$) and the transmit peak power $P_\mathrm{peak}$. 
Then, a detailed analysis is presented for the specific case of Venus in worst-case range conditions.

\begin{table}[t]
	\centering
	\caption{Maximum achievable ranges, in AU, for given PPM order and $\sigma_{\theta}$. Results found using Gaussian beam model, $R = 1/3$, $T_\mathrm{s} = 256 \, \mathrm{ns}$, $n_\mathrm{b} = 1.21\cdot 10^7\, \mathrm{phe/s}$, $P_{\mathrm{av}} = 5\,\mathrm{W}$, $P_\mathrm{out}=5$\%, $\lambda = 1064\,\mathrm{nm}$, link margin $3\,\mathrm{dB}$.}
	\footnotesize
	\begin{tabular}{l|ccccc}
		\toprule
		& \multicolumn{5}{c}{\textbf{PPM Order / Peak Laser Power [W]}} \\
		$\sigma_{\theta} [\mathrm{\upmu rad}]$ & 256 / 1600 & 64 / 400 & 32 / 200 & 16 / 100 & 4 / 25 \\
		\midrule
		1.00 & 0.113 & 0.062 & 0.046 & 0.034 & 0.019\\
		0.50 & 0.453 & 0.249 & 0.184 & 0.136 & 0.075\\
		0.35 & 0.924 & 0.507 & 0.375 & 0.277 & 0.152\\
		0.20 & 2.836 & 1.555 & 1.151 & 0.851 & 0.467\\
		0.15 & 5.037 & 2.762 & 2.045 & 1.512 & 0.830\\
		0.10 & 11.342 & 6.218 & 4.605 & 3.406 & 1.869\\
		0.05 & 45.350 & 24.864 & 18.411 & 13.617 & 7.474\\
		%0.01 & 1134.181 & 839.813 & 621.846 & 460.450 & 340.551 & 252.454 & 186.932\\
		\midrule
		$B_r\,\mathrm{[kbps]}$ & 32.33 & 97.00 & 161.66 & 258.66 & 517.32\\
		\bottomrule
	\end{tabular}
	\label{tab:MaxDistSigma}
\end{table}

% Table Max Dist iterative
% From script IterativeMaxDistance.m
\begin{table}[t]
  \centering
  \caption{Maximum achievable ranges, in AU, for given PPM order and $\theta_{\max}$. Results found using Gaussian beam model, $R = 1/3$, $T_\mathrm{s} = 256\,\mathrm{ns}$, $n_\mathrm{b} = 1.21\cdot 10^7\,\mathrm{phe/s}$, $P_{\mathrm{av}} = 5\,\mathrm{W}$, $\lambda = 1064\,\mathrm{nm}$, link margin $3\,\mathrm{dB}$.}
  \footnotesize
  \begin{tabular}{l|ccccc}
    \toprule
    & \multicolumn{5}{c}{\textbf{PPM Order / Peak Laser Power [W]}} \\
    $\theta_{\max} [\mathrm{\upmu rad}]$ & 256 / 1600 & 64 / 400 & 32 / 200 & 16 / 100 & 4 / 25 \\
    \midrule
    1.00 & 0.539 &  0.295 & 0.219 & 0.162 & 0.089\\
    0.50 & 2.155 &  1.182 & 0.875 & 0.647 & 0.355\\
    0.35 & 4.397 & 2.411 & 1.785 & 1.320 & 0.725\\
    0.20 & 13.470 & 7.385 & 5.468 & 4.044 & 2.220\\
    0.15 & 23.807 & 13.053 & 9.665 & 7.148 & 3.924\\
    0.10 & 53.880 & 29.541 & 21.874 & 16.178 & 8.880\\
    %0.05 & 215.518 & 159.582 & 118.164 & 87.495 & 64.712 & 47.972 & 35.521\\
    %0.01 & 5388.022 & 3989.600 & 2954.128 & 2187.406 & 1617.817 & 1199.304 & 888.033\\
    \midrule
    $B_r\,\mathrm{[kbps]}$ & 32.33 & 97.00 & 161.66 & 258.66 & 517.32\\
    \bottomrule
  \end{tabular}
  \label{tab:MaxDistTheta}
\end{table}

The analysis has been carried out using an efficiency equal to $-5\,\mathrm{dB}$ for both the transmit and the receive antennas, a link margin of $3$~dB (as in \cite{ITU:06}), and supplementary detection and implementation losses equal to $-4$~dB (as in \cite{BisHemPia:10}). 
The mean background noise flux is set to $n_\mathrm{b} = 1.21 \cdot 10^{7}\,\mathrm{phe/s}$, the average transmit power to $P_\mathrm{ave} = 5\,\mathrm{W}$, and the wavelength to $\lambda = 1064\,\mathrm{nm}$.
The target leg-1 decoding error probability is set to $P^{*}_\mathrm{e,1} = 9 \cdot 10^{-5}$; using $P^{*}_\mathrm{e,2} = 1 \cdot 10^{-6}$ in leg-2, this choice guarantees a total decoding error probability fulfilling $P^{*}_\mathrm{e} \le 1 \cdot 10^{-4}$ (Section~\ref{subsec:PeTot}). The outage probability of the system due to miss-pointing, when the probabilistic approach is used, is set to $5\%$; moreover, the Gaussian beam model is adopted (worst case) and the transmit and receive antennas are assumed equal in terms of diameter and pointing accuracy.
The choice of the numerical value of $n_\mathrm{b}$ deserves some explanation. 
As the maximum achievable range analysis does not refer to a specific scenario (i.e., to a deep space spacecraft probe a specific planet), a ``universal'' value of $n_\mathrm{b}$ need to be employed. 
The value $n_\mathrm{b}=1.2 \cdot 10^7\,\mathrm{phe/s}$ has been chosen because it is often suggested in literature (e.g., \cite{Biswas2020:Deep}, together with wavelength $\lambda=1064\,\mathrm{nm}$) as a ``high noise'' condition in deep space optical links.

Under this setting, we can investigate the maximum achievable range when the most powerful \ac{SCPPM} configuration \cite{CCSDS2019:142.0-B-1} is used. 
This configuration is characterized by $T_\mathrm{s} = 256\,\mathrm{ns}$ and $R = 1/3$. Table~\ref{tab:MaxDistSigma} and Table~\ref{tab:MaxDistTheta} report the maximum achievable range values, expressed in astronomic units, for the two different pointing losses estimation approaches (probabilistic and deterministic) and for different choices of the system parameters, such as the \ac{PPM} order $M$. 
For each value of $M$, the peak transmit power corresponding to an average power $P_\mathrm{ave} = 5\,\mathrm{W}$ is also shown, together with the information bit rate $B_r$ in the bottom line of the table. 
For a better interpretation of the obtained maximum achievable ranges and for the reader’s convenience, we recall that the worst-case ranges of the four considered planets are: $1.7\,\mathrm{AU}$ (Venus), 2.6 AU (Mars), 21.1 AU (Uranus), 31.3 AU (Neptune). 
The obtained results reveal how the pointing accuracy requirement is a critical parameter in the optical link design. 
This is related to the impossibility to achieve unbounded gains when pointing losses are included in the model, as seen in Section~\ref{subsec:Pointing}. 
The pointing accuracy requirement is even more critical when, as in our analysis, pointing losses are included in both the receiver and the transmitter.
Overall, we see that sub-$\mu$rad level pointing accuracy is required to establish an optical leg-1.
A specific and analysis is presented next for one of the scenarios of interest.

\begin{table}[t]
	\centering
	\caption{Example of Venus TM optical link budget with pointing losses on both the transmitter and the receiver. Maximum effective system gain.}
	\scriptsize
	\begin{tabular}{lC{1cm}C{1cm}C{1cm}}
		\toprule
		\textbf{Link Parameter} & dB &  & Units \\
		\midrule
		\rowcolor[gray]{.95} 
		\textit{Signaling and Fixed Parameter} & & & \\
		PPM Order & & 64 & \\
		Convolutional Code Rate & & 1/3 & \\
		Slot Time & & 256 & ns\\
		Guard Time & & 25 & \% \\
		Mean Noise Flux & -3.84 & 0.413 & phe/ns\\
		Mean Noise Flux per slot & & 105.63 & phe/slot \\
		$\theta_\mathrm{max}$ pointing accuracy, Gaussian Beam & & 0.35 & μrad \\
		\midrule
		\rowcolor[gray]{.95}
		\textit{Laser Transmitter} & & & \\
		Average Laser Power & 6.99 & 5.00 & W \\
		Peak Laser Power & 26.02 & 400 & W \\
		Wavelength & & 1550 & nm \\
		\midrule
		\rowcolor[gray]{.95}
		\textit{Deep Space Orbiter} & & & \\
		Far-Field Antenna Gain & 129.00 & 1.39 & m \\
		Transmitter Efficiency & -5.00 & & \\
		\midrule
		\rowcolor[gray]{.95}
		\textit{Range} & & & \\
		Space Loss & -366.46 & 1.74 & AU  \\
		\midrule
		\rowcolor[gray]{.95}
		\textit{Near Earth Orbiter} & & &\\
		Receiver Gain & 129.00 & 1.39 & m \\
		Receiver Efficiency & -5.00 & & \\
		\midrule
		\rowcolor[gray]{.95}
		\textit{Other} & & &\\
		Detection/Implementation Losses & -4.00 & & \\
		Pointing Loss & -8.45 &  & \\
		\midrule
		\rowcolor[gray]{.95}
		\textit{Link Performance} & & &\\
		Average Received Power & -123.93 & & W  \\
		Average Received Photon Flux & -25.00 & 3.16e-03 & phe/ns \\
		Minimum Average Received Power & -127.60 &  & W  \\
		Minimum Average Received Photon Flux & -28.68 & 1.36e-03 & phe/ns \\
		Link Margin & 3.67 & & \\
		FER target & & 9.00e-05 & \\
		Information Data Rate & & 0.10 & Mbps\\
		\bottomrule
	\end{tabular}
	\label{tab:LinkBudgetMars}
\end{table}

An accurate optical link budget for leg-1 has been obtained in the Venus scenario, assuming as usual the worst-case range.
It is presented in Table~\ref{tab:LinkBudgetMars}.
Regarding the mean noise flux $n_\mathrm{b}$, its value for this specific scenario has been estimate using the model in \cite{ITU:06} as reported in Section~\ref{subsec:OptLinkAnalysis}.
The link budget shows that a data rate of $100\,\mathrm{kbps}$, the same supported by a direct Venus-to-Earth link (Table~\ref{table:data_rates}) can be supported by leg-1 of a two-leg system with optical technology.  
Notably, this is possible with an optical antenna size of $1.39\,\mathrm{m}$ onboard \acp{S/C} (both the deep space one and the relay) and a pointing accuracy of $\theta_\mathrm{max} = 0.35$~μrad. 
When these requirements are met, the two-leg system (including leg-1 and leg-2) achieves the same data rate and reliability and the direct link system, using \ac{G/S} antenna sizes remarkably lower than DSA ones, as discussed above in Section~\ref{subsubsec:leg2}.

\section{System Engineering Resources for an optical link}\label{sec:sys_res_optical}
Given that optical communication in deep space is not yet a mature technology, the impact of implementing such link in leg-1 of a 2-leg data relay architecture needs to be addressed from a system resources perspective. Accordingly, mass/size estimates have been computed both for the deep-space and data relay terminals, although the former is expected to be the one subject to more stringent constraints since the data relay subsystem would not be the primary mission payload. These estimates are then compared to the mass that can be reasonably expected to be dedicated to the communication subsystem onboard the hosting spacecraft.  

Following \cite{Brown2002:System}, it is possible to allocate the communication subsystem mass as a percentage of the total dry mass depending on the spacecraft type from available historical data. For a planetary exploration spacecraft, this fraction is 6-7\%, and increases to 28\% for a communication satellite. The former can be assumed as the reference class for the deep space terminal, while the latter could be assumed as the reference for the data relay node. Note that a mass fraction of about 28\% is also corroborated by a study on a new light-TDRS \cite{Bhasin2014:Design}. In this framework of a two-leg deep space architecture, however, we cannot assume the entire mass fraction being allocated to data relay on leg-1, since data relay on leg-2 (towards Earth GS) might involve a different band, e.g. Ka. Thus, in this analysis we conservatively adopt a lower fraction, 20\%, for leg-1 data relay subsystem. 

Mass estimates for the two terminals are further split according to different assumed \ac{S/C} size ranges. For the deep-space terminal, light, medium, and heavy classes are considered, which yields to the following mass allocation to the communication subsystem (comm. mass):
\begin{itemize}
    \item Light: S/C dry mass = $400$-$600\,\mathrm{kg}$; comm. mass = $24$-$42\,\mathrm{kg}$;
    \item Medium: S/C dry mass = $1000$-$1200\,\mathrm{kg}$; comm. mass = $60$-$84\,\mathrm{kg}$;
    \item Heavy: S/C dry mass = $2000$-$2200\,\mathrm{kg}$; comm. mass = $120$-$154\,\mathrm{kg}$.
\end{itemize}
For the data-relay spacecraft, two sizes are considered which are representative of the current generation TDRS-M spacecraft and of the light-TDRS concept, yielding to:
\begin{itemize}
    \item TDRS-M: S/C dry mass = $1800\,\mathrm{kg}$; comm. mass = $504\,\mathrm{kg}$; leg-1 comm. ss mass = $360\,\mathrm{kg}$;
    \item Light-TDRS: S/C dry mass = $745\,\mathrm{kg}$; comm. mass = $208.6\,\mathrm{kg}$, leg-1 comm. mass = $149\,\mathrm{kg}$.
\end{itemize}

\subsection{Engineering Resources for an Optical Link}

According to the method in \cite{Biswas2020:Deep}, the optical transceiver mass can be expressed as the sum of:
\begin{itemize}
    \item $M_{\Opt}$: mass of the optical head of the telescope, which shows a dependency on the telescope aperture diameter;
	\item $M_{\las}$: mass of the transmitter equipment, which shows a dependency on the laser average power.
\end{itemize}

The empirical equations for computing $M_{\Opt}$ and $M_{\las}$ as a function of the optics diameter and average power are \cite{Biswas2020:Deep}:
\begin{align}
M_{\Opt} \left[\mathrm{kg}\right] &= K\cdot D^{2.57} \\
M_{\las} \left[\mathrm{kg}\right] &= 1.152\cdot P_{\Opt}+3.168 \notag
\end{align}
where $K=0.00181$, $D$ is the telescope diameter in centimeters, and $P_{\mathrm{avg}}$ is the laser average power.
Allocating the mass for the deep space optical communication segment according to the above estimates for the three \ac{S/C} classes, and assuming $P_{\mathrm{avg}}=5\,\mathrm{W}$, the maximum diameters of the optical assembly would be:
\begin{itemize}
	\item Light: $D=34$-$42\,\mathrm{cm}$;
	\item Medium: $D=54$-$63\,\mathrm{cm}$; 
	\item Heavy: $D=73$-$81\,\mathrm{cm}$.
\end{itemize}
Following the same logic for the data relay, the equivalent diameters would instead be:
\begin{itemize}
	\item TDRS-M: $D=114\,\mathrm{cm}$;
	\item Light-TDRS: $D=80\,\mathrm{cm}$.
\end{itemize}
   
In all cases, with the same mass allocation considered for traditional RF communication segments, the maximum diameter achievable for the optical terminal is lower than the optimal diameter estimated in Section~\ref{subsubsec:optical_results}, equal to $139\,\mathrm{cm}$. 
Accordingly, the actual gain would be lower. 
The difference between optimum diameter and achievable one is smaller for the data relay \ac{S/C}, and it is reasonable to assume, in this case, a greater engineering effort to accommodate a more performing communication equipment. 
Although the optical antenna diameter is comparable with the diameter of traditional parabolic dish antennas commonly employed in deep space missions, the mass associated with the optical diameter results considerably higher than that of traditional RF antennas. 
This leads to the recommendation of investing in the development of optical antenna designs optimized for reducing the gain/mass ratio.

\section{Conclusions and Perspectives}\label{sec:conclusions_perspectives}

In this work, we analyzed two-leg deep space relay architectures for assessing their potential advantages over classical direct RF links for data downlink. 
The problem was tackled from several perspectives. 
A mathematical framework for analysis of two-leg deep space relay communication systems was first developed, addressing both the transparent and the regenerative relay cases. 
In the former case, the \ac{EHF} band was considered in leg-1 (the interplanetary link), while in the latter case the analysis included both optical and \ac{EHF} technologies in leg-1, as well as classical \ac{RF} frequencies in leg-2.
Different two-leg orbital architectures were assessed, assuming the relay to be placed whether on LEO, GEO or at a Lagrange point of the Earth-Sun system, and the deep space probe orbiting at an inner or outer planet. 
The various combinations were then explored in terms of range, complexity and  expected outage periods due to the presence of the Sun in the vicinity of the optical terminals' \ac{FOV}s. 
Our results highlighted that two-leg deep space relay systems, when used with regenerative relays, can provide advantages over direct link systems, provided that specific values of the system parameters can be supported. 

The main conclusions may be summarized as follows:
\begin{itemize}
    \item A two-leg regenerative relay system with optical frequencies in leg-1 can achieve the same data rates achieved in direct links in the Venus scenario, provided a pointing accuracy of 0.35 μrad can be guaranteed onboard \acp{S/C} (deep space probe and relay), together with an average transmit power of $5\,\mathrm{W}$, a peak transmit power of $400\,\mathrm{W}$, and an optical antenna size in the order of $1.4\,\mathrm{m}$. 
    
    Although a detailed optical link budget was shown in Section~\ref{subsubsec:optical_results} only for the Venus case, the analysis can be repeated for the other scenarios. In the Mars case we obtained the same parameters and pointing requirement.
    Regarding Uranus and Neptune, more tightening pointing accuracy constraints have been obtained, as low as 0.1 μrad, together with an average transmit power of $5\,\mathrm{W}$, a peak transmit power of $400\,\mathrm{W}$, and an optical antenna size of about $5\,\mathrm{m}$. 
    \item Two-leg regenerative relay systems operating in EHF band in leg-1 can achieve the same data rates achieved in direct links, provided a pointing accuracy of $0.01$° can be guaranteed onboard \acp{S/C} (in all scenarios), with a leg-1 frequency $f_1 \approx 64$ GHz, and an antenna size in the order of $5\,\mathrm{m}$.
    \item In both cases (optical and \ac{EHF}), due to the large margin available in leg-2, antenna sizes smaller than the typical DSA ones can be used on ground. For a GEO relay, antenna diameters of $2.4\,\mathrm{m}$ (corresponding, for example, to ESA REDU-3) can be employed in all scenarios with a frequency $f_2 = 27\,\mathrm{GHz}$. 
    For a Lagrange L4/L5 relay and a Ka-band relay-to-ground link, antenna diameters of $2.4\,\mathrm{m}$ can be employed in the Uranus and Neptune scenarios, while larger antenna sizes, yet smaller than DSA ones, are necessary in the Venus and Mars ones due to the higher data rate. 
    For example, with a concatenated coding scheme, diameters in the order of $7.7\,\mathrm{m}$ and $8.4\,\mathrm{m}$ are necessary for the Venus and Mars scenarios, respectively.
	\item In both cases (optical and \ac{EHF}), the two-leg system bottleneck is represented by leg-1. This is mainly due to pointing losses severely affecting both the transmitter and the receiver. In contrast, the available analyses for direct link systems, RF or optical, include pointing losses on the spacecraft side but not on the ground side.
	\item In case of optical frequencies, different planetary targets may lead to different expected mission coverage in downlink, depending on the assumed minimum offset angle the optical terminal is designed to operate at. Stringent design constraints are expected on the optical terminal of a deep space probe at Uranus or Neptune, as it must be capable to operate at an \ac{SPD} angle always smaller than 0.5°. 
	\item A two-leg architecture with a relay spacecraft in L4/L5 would take advantage of a larger angular separation from the Earth, so that when either L4 or L5 are at low \ac{SDP} angles, ground-based deep space antennas on the Earth may be used to send telecommands to the deep space probe, thereby virtually eliminating any outage due to occultations.
	\item With the same mass allocation considered for traditional RF communication segments, the maximum diameter obtainable for the optical terminal onboard the deep space spacecraft is smaller than that required by the optimal effective system gain. It is thus recommended to seek for novel optical antenna designs featuring increased gain/mass ratios.
\end{itemize}

\begin{table*}[t]
	\centering
	\caption{Summary of Link Analysis.}
	\begin{tabular}{l|C{3cm}C{3cm}C{3cm}}
		\toprule
		\rowcolor[gray]{.95} 
		\textbf{Configuration} & \textbf{Direct Link} & \makecell{\textbf{Two-Leg}\\ \textbf{Optical Leg-1}} &  \makecell{\textbf{Two-Leg}\\\textbf{EHF Leg-1}} \\
		\midrule
		\textbf{Tx Power} & $65$~W & \makecell{$5$~W (average) \\ $400$~W (peak)}& $65$~W \\ \hline
		\textbf{Leg-1 Band} & $8.42$ GHz & $1550$~nm & $63.9$~GHz\\ \hline
		\textbf{Leg-2 Band} & $-$ & \makecell{$27$ GHz (GEO/L1)\\$32.4$ GHz (L4/L5)} & \makecell{$27$ GHz (GEO/L1)\\$32.4$ GHz (L4/L5)} \\ \hline
		\textbf{Relay Rx Antenna} & $-$ & 1.39 m & 5 m \\
		\midrule
		\rowcolor[gray]{.95}
		\textbf{Venus (V)} -- 1.73 AU & & & \\
		\textbf{Deep Space S/C Antenna} & $1.3$ m& $1.39$ m& 1.3 m\\
		\textbf{Data Rate} & 100 kbps & 100 kbps & 100 kbps\\
		\textbf{S/C Pointing Accuracy} & 0.05° & 0.35 μrad & 0.01°\\
		\rowcolor[gray]{.95}
		\textbf{Mars (M)} -- 2.68 AU & & & \\
		\textbf{Deep Space S/C Antenna} & 2.2 m& $1.39$ m& 2.2 m\\
		\textbf{Data Rate} & 120 kbps& 100 kbps& 120 kbps\\
		\textbf{S/C Pointing Accuracy} & 0.05° & 0.35 μrad & 0.01°\\
		\rowcolor[gray]{.95}
		\textbf{Uranus (U)} -- 21.1 AU & & & \\
		\textbf{Deep Space S/C Antenna} & 3 m& $5$ m& 3 m\\
		\textbf{Data Rate} & 3.15 kbps& 100 kbps& 3.15 kbps\\
		\textbf{S/C Pointing Accuracy} & 0.05° & $0.1$~μrad & 0.01°\\
		\rowcolor[gray]{.95}
		\textbf{Neptune (N)} -- 31.3 AU & & & \\
		\textbf{Deep Space S/C Antenna} & 3 m& $5$ m& 3 m\\
		\textbf{Data Rate} &  1.20 kbps& 100 kbps& 1.20 kbps\\
		\textbf{S/C Pointing Accuracy} & 0.05° & $0.1$~μrad & 0.01°\\
		\midrule 
		\rowcolor[gray]{.95}
		 & & & \\
		\textbf{Ground Antenna} & 35 m & \makecell{2.4 m (GEO/L1-all) \\ 7.7 m (L4/L5-all)} & \makecell{2.4 m (GEO/L1-all) \\ 2.4 m (L4/L5-U/N) \\ 7.7 m (L4/L5-V) \\ 8.4 m (L4/L5-M)} \\
		\bottomrule
	\end{tabular}
	\label{tab:SummaryLinkAnalysis}
\end{table*}

A summary of the most significant link analysis results is shown in Table~\ref{tab:SummaryLinkAnalysis}, for the regenerative relay case. 
In the table, the two-leg architecture based on \ac{EHF} in leg-1 is compared with a direct X-band link for the same information bit rate. 
In case of optical technology used in leg-1, instead, the \ac{SCPPM} parameters $M=64$, $R=1/3$, and $T_{\mathrm{s}}=256\,\mathrm{ns}$ are the same in all scenarios, resulting in the same data rate.
These parameters are addressed in the appendix.

Since both optical and \ac{EHF} technologies can guarantee similar advantages in leg-1, the choice of the system architecture is mainly dictated by the capability of current systems to support the above-mentioned set of parameters; out of them, the most critical ones seem to be represented by the required pointing accuracy levels in the optical case and by the relay receive antenna size in the RF one. 
The choice of the specific leg-1 technology may change depending on the scenario.
For example, should the optical pointing requirement for Mars and Venus be feasible, while the one for Uranus and Neptune be too tight, an optical system may be chosen in the former scenarios and an RF one for the latter ones.

We close the paper by pointing out that the use of the optical technology in leg-2 has not been addressed in this work for several reasons, including: (1) The significant impact of the atmospheric effects, potentially quite severe especially at low elevation angles; (2) Problems arising in the spacecraft pointing accuracy, as an optical uplink beacon would be significantly affected (much more than the downlink) by scintillation and beam wandering, again due to the Earth atmosphere; (3) The fact that the technology for L4/L5-to-ground optical links is not as mature and consolidated as it is for inter-satellite links or as the \ac{RF} one \cite{Sodnik2017:Deep}.
It is however undoubted that the technological development of optical communications is currently gaining a momentum, as witnessed by the intensified standardization activities within the \ac{CCSDS} and by a set of in-orbit demonstrators and missions from several space agencies, including ESA HydRON \cite{Perdigues2021:HYDRON}, NASA TBIRD \cite{Schieler2019:Demonstration}, and DLR Osiris \cite{Fuchs2018:Update}.
The analysis of \ac{RF}-optical or optical-optical two-leg deep space relay systems is therefore an important challenging direction of further investigation.

\appendices

\section{SCPPM Coded Modulation}\label{appendix:SCPPM}

The SCPPM encoding scheme accepts as input information blocks with configurable size and generates a sequence of coded \ac{PPM} symbols which form an SCPPM codeword \cite{CCSDS2019:142.0-B-1,Moision2005:Coded}. Depending on the chosen code rate of the convolutional encoder, whose admitted values are $1/3$, $1/2$, and $2/3$, the information bits are sliced into blocks whose sizes are reported in Table~\ref{table:SCPPM}. A $32$-bit cyclic redundancy check (CRC) code, used for block error detection, is attached to each information block. Then, before the block is passed to the SCPPM encoder, two 0 bits are appended to it (termination bits) in order to let the convolutional encoder terminate in the zero state before the next block is processed. The input block so generated is denoted by $\bm{u}$. Its admitted sizes are shown in the third column of Table~\ref{table:SCPPM}.

The SCPPM encoder is composed of an outer convolutional encoder, a bit interleaver, and an accumulator-PPM (APPM) referred to as the “inner encoder”. The convolutional encoder is described by the generator polynomials 
\begin{align}
& g^{(1)} (D) = 1+D^2 \notag \\        
& g^{(2)} (D) = 1+D+D^2 \\
& g^{(3)} (D) = 1+D+D^2 \notag
\end{align}
along with the puncturing patterns specified in \cite{CCSDS2019:142.0-B-1}. 
The three code rates require three different lengths for the input block $\bm{u}$, as previously described, to generate a fixed-length output of $15120$ bit. 
The advantage of adopting a fixed length output (i.e., coded) block is the possibility to implement a single interleaver, specified by the function
\begin{align}
\pi(j) = 11 j + 210 j^2 \,\, \mathrm{mod}\,\, 15120
\end{align}
where $j$ is the index of the input bit and $\pi(j)$ is the corresponding output index. 
The scrambled bits are then processed by the APPM (inner encoder). 
The bits are first processed by a rate-1 accumulator and then they are divided into sub-blocks of $m = \log_2 M$ bit each, where $M$ is the PPM order, $M \in \{4,8,16,32,64,128,256\}$. 
Each sub-block of $m$ bits is mapped onto a \ac{PPM} symbol, yielding $15120 / \log_2 M$ PPM symbols per input block. 
A Gray or anti-Gray bit mapper may be introduced between slicing into sub-blocks and PPM modulation. 
The SCPPM standard also imposes a $25\%$ guard time between the \ac{PPM} symbols, where an $M$-\ac{PPM} symbol is represented as a sequence of $M$ time slots, of which only one is pulsed (the pulsed slot depends on the value of the corresponding $m$ bits). 
Typical values for the slot time $T_\mathrm{s}$ are in the range $0.5\,\mathrm{ns}$ - $256\,\mathrm{ns}$. Hence, each \ac{PPM} symbol has a time duration equal to $M T_\mathrm{s}$; the guard time duration is $M T_\mathrm{s} /4$.

\begin{table}[!t]
\begin{center}
\caption{Input block sizes for the three possible SCPPM code rates.}\label{table:SCPPM}
\footnotesize{
\begin{tabular}{ccc}
\toprule
 {\textbf{Convolutional}} & {\textbf{Information block}} & {\textbf{Input block size }}\\
 {\textbf{code rate}} & {\textbf{size 
(bits)}} & {\textbf{to SCPPM encoder (bits)}}\\
 \toprule
$1/3$ & $5006$ & $5040$\\
$1/2$ & $7526$ & $7560$\\
$2/3$ & $10046$ & $10080$\\
\bottomrule
\end{tabular}}
\end{center}
\end{table}

An efficient turbo-like iterative decoding scheme can be used to decode \ac{SCPPM} codes. 
This scheme, based on two soft-input soft-output (SISO) module exchanging information in an iterative fashion, allows achieving a performance that approaches the one obtained by optimum maximum a posteriori decoding.
A ``logical'' $M$-\ac{PPM} symbol can be thought as a vector of size $M$ in which all elements are null except one element, corresponding to the pulsed slot, which is equal to one. When this symbol is transmitted by a laser system, we can represent the received symbol as a vector $\bm{y}$, again of size $M$, each element of which is a nonnegative integer representing the number of photo-detected photons. 
Considering a Poisson PPM channel, a typical optical channel model, the conditional probabilities to detect $k$ photons in a given slot, conditioned to the slot being empty (``0'') or pulsed (``1''), are given by
\begin{align}
p_{Y | C} (k | 1) &= \frac{(n_\mathrm{s}+n_{\mathrm{b}} )^k  e^{-( n_\mathrm{s}+n_\mathrm{b} ) }}{k!} \\
p_{Y | C} (k | 0) &= \frac{n_\mathrm{b}^k  e^{-n_\mathrm{b}}}{k!}                    
\end{align}
where $n_\mathrm{b}$ is the average number of noise photons per slot and $n_\mathrm{s}$ is the average number of received signal photons in a pulsed slot. 

\begin{figure}
    \centering
    \resizebox{0.45\textwidth}{!}{
        % CURVA PRESTAZIONI	
% This file was created by matlab2tikz.
%
%The latest updates can be retrieved from
%  http://www.mathworks.com/matlabcentral/fileexchange/22022-matlab2tikz-matlab2tikz
%where you can also make suggestions and rate matlab2tikz.
%
\definecolor{newBlue}{rgb}{0.1804, 0.5255, 0.7569}%
\definecolor{graphRed}{rgb}{0.76078,0.13725,0.13725}%

\begin{tikzpicture}
	
	\begin{axis}[%
		width=4.521in,
		height=3.552in,
		at={(0.758in,0.495in)},
		scale only axis,
		xmin=-35,
		xmax=-15,
		xlabel style={font=\color{white!15!black}},
		xlabel={$n_s/(MT_{s})$, dB photons/ns},
		ymode=log,
		ymin=1e-06,
		ymax=1,
		yminorticks=true,
		ylabel style={font=\color{white!15!black}},
		ylabel={FER},
		axis background/.style={fill=white},
		xmajorgrids,
		ymajorgrids,
		yminorgrids,
		legend style={at={(0.97,0.03)}, anchor=south east, legend cell align=left, align=left, draw=white!15!black}
		]
		\addplot [color=black, line width=0.8pt, mark=star, mark size=2.5, mark options={solid, black}]
		table[row sep=crcr]{%
			-16.4481197174891	0.762\\
			-16.3886263923571	0.396\\
			-16.3299370557161	0.169326856349757\\
			-16.2720302669072	0.0365983655098898\\
			-16.2148854316629	0.00529761978438192\\
			-16.158482758136	0.000632045466660236\\
			-16.1028032157465	2.4e-05\\
			-16.0478284966344	1.6e-05\\
		};
		\addlegendentry{$M = 4$, $R = 1/2$}
		
		%\addplot [color=graphRed, draw=none, mark=o, mark options={solid, graphRed}, forget plot]
		%  table[row sep=crcr]{%
			%-16.1028032157465	2.4e-05\\
			%};
		%\addplot [color=graphRed, draw=none, mark=o, mark options={solid, graphRed}, forget plot]
		%  table[row sep=crcr]{%
			%-16.0478284966344	1.6e-05\\
			%};
		\addplot [color=black, line width=0.8pt, mark=diamond, mark size=2.5, mark options={solid, black}]
		table[row sep=crcr]{%
			-19.5897761000296	1\\
			-19.4395912013524	1\\
			-19.289814587121	1\\
			-19.1397402800811	0.726\\
			-18.9895677657518	0.0904625460499386\\
			-18.8397369654487	0.000842\\
		};
		\addlegendentry{$M = 8$, $R = 1/2$}
		
		\addplot [color=black, line width=0.8pt, mark=triangle, mark size=2.5, mark options={solid, rotate=180, black}]
		table[row sep=crcr]{%
			-22.0411998265592	0.998\\
			-21.9738659999696	0.992\\
			-21.9075602109794	0.944\\
			-21.8422515393899	0.724\\
			-21.7779104393358	0.376\\
			-21.7145086590256	0.14682249817385\\
			-21.6520191662556	0.029147216213139\\
			-21.5904160792074	0.00386681753788858\\
			-21.5296746020854	0.000241\\
			-21.4697709651936	0\\
		};
		\addlegendentry{$M = 16$, $R = 1/2$}
		
		%\addplot [color=graphRed, draw=none, mark=o, mark options={solid, graphRed}, forget plot]
		%  table[row sep=crcr]{%
			%-21.4697709651936	0\\
			%};
		\addplot [color=black, line width=0.8pt, mark=triangle, mark size=2.5, mark options={solid, black}]
		table[row sep=crcr]{%
			-24.7882103959756	1\\
			-24.7248086156654	0.996\\
			-24.6623191228954	0.996\\
			-24.6007160358472	0.932\\
			-24.5399745587252	0.744\\
			-24.4800709218334	0.482\\
			-24.4209823257282	0.21151832460733\\
			-24.3626868891209	0.0589367332210729\\
			-24.30516360023	0.0107163178858191\\
			-24.2483922713131	0.00125901206625587\\
			-24.1923534961331	0.000135\\
			-24.1370286101335	1.4e-05\\
			-24.0823996531185	0\\
		};
		\addlegendentry{$M = 32$, $R = 1/2$}
		
		%\addplot [color=graphRed, draw=none, mark=o, mark options={solid, graphRed}, forget plot]
		%  table[row sep=crcr]{%
			%-24.1370286101335	1.4e-05\\
			%};
		%\addplot [color=graphRed, draw=none, mark=o, mark options={solid, graphRed}, forget plot]
		%  table[row sep=crcr]{%
			%-24.0823996531185	0\\
			%};
		\addplot [color=black, line width=0.8pt, mark=asterisk, mark size=2.5, mark options={solid, black}]
		table[row sep=crcr]{%
			-27.1897551484739	0.696\\
			-27.0897040111959	0.219195849546044\\
			-27.0396930456339	0.0760598503740648\\
			-26.9897000433602	0.0227582182454775\\
			-26.8897084324679	0.000834447903227715\\
			-26.8690345950794	0.000191018387502063\\
		};
		\addlegendentry{$M = 64$, $R = 1/2$}
		
		\addplot [color=black, line width=0.8pt, mark=o, mark size=2.5, mark options={solid, black}]
		table[row sep=crcr]{%
			-30.2129534094128	1\\
			%-30.1029995663981	1\\
			-29.9957609124804	0.99\\
			-29.8911065756987	0.836\\
			-29.7889149238819	0.424\\
			-29.6890727148159	0.0792741165234002\\
			-29.5914743419243	0.00573134012994559\\
			-29.496021162862	0.000176\\
			-29.4771797241165	0.000110629906270812\\
			-29.4026209003206	0\\
			-29.3111871059219	0\\
		};
		\addlegendentry{$M = 128$, $R = 1/2$}
		
		%\addplot [color=graphRed, draw=none, mark=o, mark options={solid, graphRed}, forget plot]
		%  table[row sep=crcr]{%
			%-29.4026209003206	0\\
			%};
		%\addplot [color=graphRed, draw=none, mark=o, mark options={solid, graphRed}, forget plot]
		%  table[row sep=crcr]{%
			%-29.3111871059219	0\\
			%};
		\addplot [color=black, line width=0.8pt, mark=square, mark size=2.5, mark options={solid, black}]
		table[row sep=crcr]{%
			-32.5063211195018	0.742\\
			-32.4129208569604	0.336\\
			-32.3214870625617	0.0679702048417132\\
			-32.2319386360324	0.00577367205542725\\
			-32.1441993929574	9.8e-05\\
			-32.1268622648674	3.13333333333333e-05\\
			-32.0581976753382	0\\
			-31.9738659999696	0\\
			-31.8911407403097	0\\
		};
		\addlegendentry{$M = 256$, $R = 1/2$}
		
		%\addplot [color=blue, draw=none, mark=o, mark options={solid, blue}, forget plot]
		%  table[row sep=crcr]{%
			%-32.1441993929574	9.8e-05\\
			%};
		%\addplot [color=blue, draw=none, mark=o, mark options={solid, blue}, forget plot]
		%  table[row sep=crcr]{%
			%-32.1268622648674	3.13333333333333e-05\\
			%};
		%\addplot [color=graphRed, draw=none, mark=o, mark options={solid, graphRed}, forget plot]
		%  table[row sep=crcr]{%
			%-32.0581976753382	0\\
			%};
		%\addplot [color=graphRed, draw=none, mark=o, mark options={solid, graphRed}, forget plot]
		%  table[row sep=crcr]{%
			%-31.9738659999696	0\\
			%};
		%\addplot [color=graphRed, draw=none, mark=o, mark options={solid, graphRed}, forget plot]
		%  table[row sep=crcr]{%
			%-31.8911407403097	0\\
			%};
		\addplot [color=graphRed, line width=0.8pt, mark=asterisk, mark size=2.5, mark options={solid, graphRed}]
		table[row sep=crcr]{%
			-28.9635660433298	1\\
			-28.6417192096157	0.934\\
			%-28.6108084398443	0.874\\
			%-28.595435167795	0.84\\
			%-28.5801161225676	0.766\\
			-28.5648509229528	0.732\\
			%-28.5496391917475	0.664\\
			-28.5193746454456	0.538\\
			%-28.4893195440484	0.428\\
			-28.3420869758413	0.0346089850249584\\
			-28.3132300934664	0.0151286921212915\\
			-28.2845636869504	0.00680272108843537\\
			-28.2560852582221	0.00304626795384978\\
			-28.2277923580335	0.00129281738283123\\
			-28.1996825846952	0.000235280500843879\\
		};
		\addlegendentry{$M = 64$, $R = 1/3$}
		
		\addplot [color=graphRed, line width=0.8pt, mark=square, mark size=2.5, mark options={solid, graphRed}]
		table[row sep=crcr]{%
			-33.7431384483898	0.436\\
			-33.6932189928148	0.265709156193896\\
			-33.6438668160597	0.107313738892686\\
			-33.5950691698788	0.0403536613012922\\
			-33.5468137309859	0.0145733596170093\\
			-33.4990885823734	0.00345425794863135\\
			-33.4518821956476	0.00074\\
			-33.4051834143128	0.000127569026499389\\
			-33.3820209870409	6.36363636363636e-05\\
			-33.3360634701495	2e-05\\
		};
		\addlegendentry{$M = 256$, $R = 1/3$}
		
		%\addplot [color=blue, draw=none, mark=o, mark options={solid, blue}, forget plot]
		%  table[row sep=crcr]{%
			%-33.4518821956476	0.00074\\
			%};
		%\addplot [color=blue, draw=none, mark=o, mark options={solid, blue}, forget plot]
		%  table[row sep=crcr]{%
			%-33.3820209870409	6.36363636363636e-05\\
			%};
		%\addplot [color=graphRed, draw=none, mark=o, mark options={solid, graphRed}, forget plot]
		%  table[row sep=crcr]{%
			%-33.3360634701495	2e-05\\
			%};
		\addplot [color=newBlue, line width=0.8pt, mark=square, mark size=2.5, mark options={solid, newBlue}]
		table[row sep=crcr]{%
			-31.4996195006882	0.986\\
			-31.4253793198961	0.914\\
			-31.3523869324811	0.7\\
			-31.2806010862099	0.338\\
			-31.209982541335	0.0825598436736688\\
			-31.1404939417817	0.0124339976153977\\
			-31.0720996964787	0.00129500002452652\\
			-31.004765869889	7e-06\\
		};
		\addlegendentry{$M = 256$, $R = 2/3$}
		
		%\addplot [color=graphRed, draw=none, mark=o, mark options={solid, graphRed}, forget plot]
		%  table[row sep=crcr]{%
			%-31.004765869889	7e-06\\
			%};
		
		\pgfplotsset{legend style={font=\tiny}}

	\end{axis}
	
\end{tikzpicture}%
    }
    \caption{Frame error rate of \ac{SCPPM} codes assuming $n_\mathrm{b} = 0.2$~phe/slot, $T_\mathrm{s} = 16$~ns, and Gray mapping.}
  \label{fig:SCPPMperformanceGraph}
\end{figure}
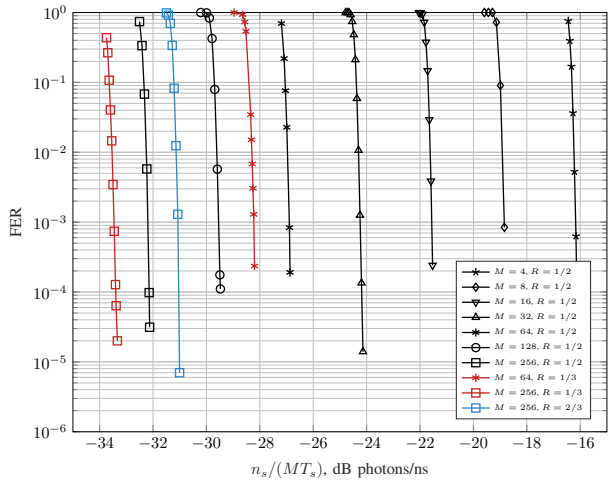

Fig.~\ref{fig:SCPPMperformanceGraph} shows the performance of SCPPM codes simulated with different values of the PPM order $M$ and code rate; all curves are relevant to a slot time $T_\mathrm{s}=16$ ns, $n_\mathrm{b}=0.2$ phe/slot, and 25 decoding iterations per codeword. As we can observe, for given $T_\mathrm{s}$, $n_\mathrm{b}$ and convolutional code rate, the larger is the PPM order and the better is the SCPPM performance; the scheme, in fact, requires a smaller average number of signal photons per nanosecond to achieve a target FER. Interestingly, the gap between different simulated curves exhibits an almost systematic behavior. More specifically, doubling $M$ turns into a coding gain of approximately $2.5$ dB. Our simulations also confirm that, as expected, for a given order M, the SCPPM scheme with code rate $1/3$ (red curves, for $M=64$ and $M=256$) achieves the best performance, followed by the one with code rate $1/2$ (black curves) and finally by the one with code rate $2/3$ (blue curve for $M=256$).

\section*{Acknowledgment}

The authors would like to thank Daniel Arapoglou (European Space Agency) for useful discussions.

\end{document}